\documentclass{article} 
\usepackage{iclr2021_conference,times}

\usepackage{hyperref}
\usepackage{url}

\usepackage{amssymb,amsmath,amsthm,enumerate,threeparttable,bm,subfigure,graphicx}
\usepackage{algorithm, algorithmic}
\usepackage{booktabs}
\usepackage{multirow}
\usepackage{color}
\usepackage{makecell}
\usepackage{float}

\newcommand{\ie}{\textit{i.e.}}
\newcommand{\eg}{\textit{e.g.}}

\long\def\comment#1{}
\newtheorem{defn}{Definition}

\title{Rethinking the Trigger of Backdoor Attack}

\author{Yiming Li$^{1}$, Tongqing Zhai$^{1}$, Baoyuan Wu$^{2,3,}$\thanks{Correspondence to: Baoyuan Wu (\href{mailto:wubaoyuan@cuhk.edu.cn}{wubaoyuan@cuhk.edu.cn}), Shu-Tao Xia (\href{mailto:xiast@sz.tsinghua.edu.cn}{xiast@sz.tsinghua.edu.cn}).} , Yong Jiang$^{1}$, Zhifeng Li$^{4}$, Shu-Tao Xia$^{1,*}$\\
$^{1}$Tsinghua Shenzhen International Graduate School, Tsinghua University, Shenzhen, China\\
$^2$School of Data Science, The Chinese University of Hong Kong, Shenzhen, China \\
$^3$Secure Computing Lab of Big Data, Shenzhen Research Institute of Big Data, Shenzhen, China\\
$^{4}$Tencent AI Lab, Shenzhen, China
}

\iclrfinalcopy 
\begin{document}

\maketitle

\begin{abstract}
Backdoor attack intends to inject hidden backdoor into the deep neural networks (DNNs), such that the prediction of the infected model will be maliciously changed if the hidden backdoor is activated by the attacker-defined trigger, while it performs well on benign samples. Currently, most of existing backdoor attacks adopted the setting of \emph{static} trigger, $i.e.,$ triggers across the training and testing images follow the same appearance and are located in the same area. In this paper, we revisit this attack paradigm by analyzing the characteristics of the static trigger. We demonstrate that such an attack paradigm is vulnerable when the trigger in testing images is not consistent with the one used for training. We further explore how to utilize this property for backdoor defense, and discuss how to alleviate such vulnerability of existing attacks. 

\end{abstract}

\vspace{-1em}
\section{Introduction}
\vspace{-0.8em}

Deep neural networks (DNNs) have demonstrated their superior performance in a variety of applications. 
However, DNNs have been proved to be unstable that the small perturbation on the input may lead to a significant change in the output, which raises serious security concerns. 
For example, given one trained DNN model and one benign sample, the malicious perturbation could be optimized to encourage that the perturbed sample will be misclassified, while the perturbation is imperceptible to human eyes. It is dubbed {\it adversarial attack}, which happens in the inference stage \citep{madry2017,fan2020sparse,bai2020targeted}.

In contrast, some recent studies showed that some regular (\ie, non-optimized) perturbations (\eg, the local patch stamped on the right-bottom corner of the image) could also mislead DNNs, through influencing the model weights in the training process \citep{liu2020survey,li2020backdoor,gao2020backdoor}. 
It is called as {\it backdoor attack}\footnote{Backdoor attack is also commonly called the `Trojan attack', such as in \citep{liu2017trojaning,ding2019trojan,chen2019deepinspect}. In this paper, `backdoor attack' refers specifically to attack methods that modify the training samples to create the backdoor, and we only focus on the image classification.}. Specifically, some training samples are modified by adding the trigger (\eg, the local patch). These modified samples with attacker-specified target labels, together with 
benign training samples, are fed into the DNN model for training. Consequently, the trained DNN model performs well on benign testing samples, similarly with the model trained using only benign samples; however, if the same trigger used in training is added onto a testing sample, then its prediction will be changed to the target label specified by the attacker. The backdoor attack could happen in the scenario that the training process is inaccessible or out of control by the user. 
Since the infected DNN model performs normally on benign samples, the user is difficult to realize the existence of the backdoor; even if the trigger is present, since it is usually just a regular local patch or even invisible, it is difficult for the user to identify the reason of the incorrect prediction. Hence, the insidious backdoor attack is a serious threat to the practical application of DNNs.

\comment{
\begin{figure}[ht]
 \centering
 \includegraphics[width=0.47\textwidth]{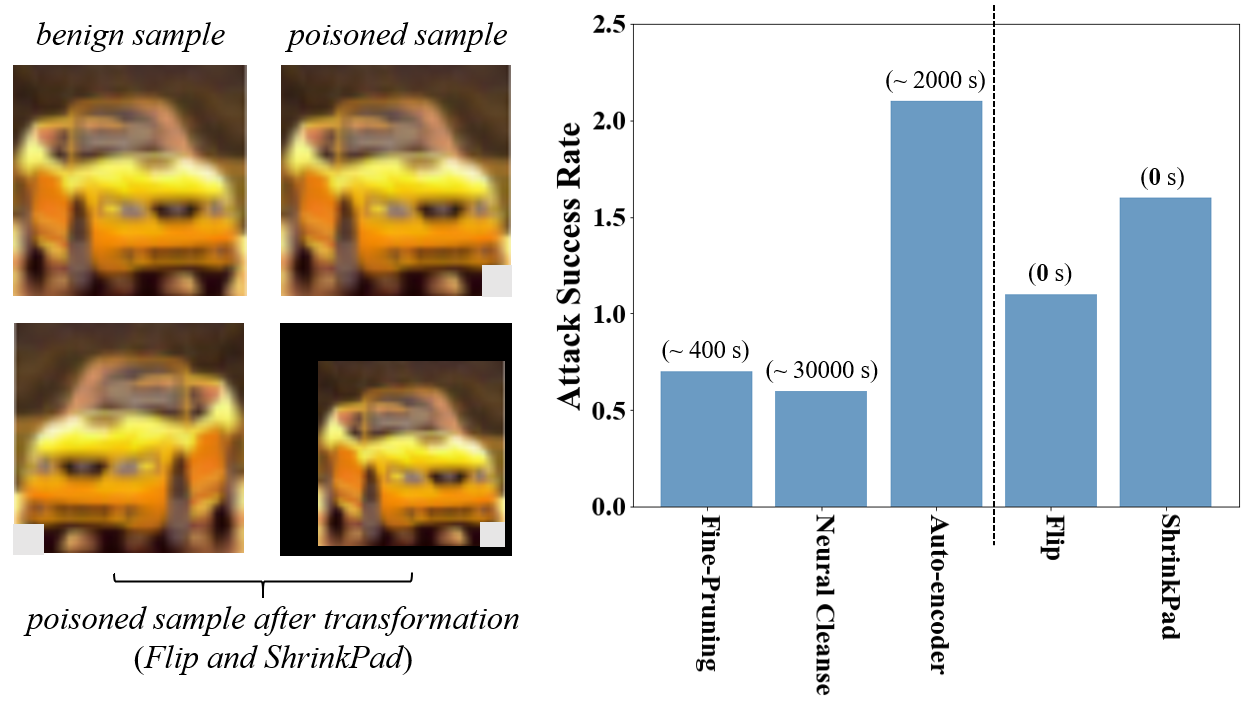}
 \caption{The comparison between different existing defenses and the proposed transformation-based defenses from the aspect of average training time and attack success rate under BadNets attack. We demonstrate two simple transformations, including flipping (Flip) and padding after shrinking (ShrinkPad) for the defense. In this example, the trigger is a 4-pixels gray square on the bottom right corner. }
 \label{vulnerability}
\end{figure}
}

Many backdoor attacks have been proposed through designing different types of triggers \citep{gu2017badnets, chen2017targeted,turner2019label,zhao2020clean,li2020backdoor1, li2020invisible}. It is interesting to find that most of existing works adopted the setting of \emph{static} trigger, where the triggers across the training and testing images are located in the same area and have the same appearance, to the best of our knowledge. 
However, the user may modify the testing images before prediction, such that the trigger's location and appearance could be changed. It raises an intriguing question: 

{\it When the trigger in the attacked testing image is different from that used in training, can it still activate the hidden backdoor?}

To answer this question, we explore the impacts of two basic characteristics of the backdoor trigger, including \emph{location} and \emph{appearance}. 
As shown in later experiments, we demonstrate that if the location or appearance of the trigger is slightly changed, then the attack performance may degrade sharply. It reveals that the backdoor attack with static trigger pattern may be non-robust to the change of trigger. The above observation inspires two further questions:

{\bf (1)} {\it Can we utilize this non-robustness to defend existing backdoor attacks?} {\bf (2)} {\it How to enhance the performance of existing backdoor attacks, such that they are robust to the change of trigger?}

In this work, we propose a simple yet effective defense method towards attacks with static trigger in which the testing sample is spatially transformed (\eg, flipping or scaling) before the prediction. The spatial transformation on the whole image is a feasible approach to change the trigger's location and appearance, which may fail to activate the hidden backdoor in the infected DNN model. 
Furthermore, we also propose to enhance the transformation robustness of the attack that all poisoned images will be randomly transformed before feeding into the training process. The proposed method is equivalent to adding a preprocessing step on the poisoned images. 
This attack enhancement could be naturally combined with any backdoor attack method. Consequently, the attack's robustness to the change of trigger is significantly enhanced, and the attack can evade the proposed transformation-based defense. 
Besides, we demonstrate the connection between the proposed attack enhancement and the physical attack, and it explains that the enhanced attack could still succeed in physical scenarios, while the standard backdoor attacks with the static trigger will fail. Moreover, we present the visualization by utilizing the \emph{saliency map} \citep{simonyan2013deep} and the \emph{critical data routing path} \citep{wang2018cdrp} of images, under the standard backdoor attack and the enhanced backdoor attack, to further understand their differences.

The main contributions of this work are three-fold: \textbf{(1)} We demonstrate that the location and appearance of the backdoor trigger have crucial impacts on activating the backdoor. \textbf{(2)} We verify that attacks with the static trigger pattern are transformation vulnerable, which inspires a simple yet effective defense. \textbf{(3)} We propose an effective method to enhance the robustness of existing attacks against the change of trigger, and connect the proposed enhancement with the physical attack. 

\comment{
\begin{itemize}
    \item We demonstrate that the location and the appearance of the backdoor trigger have crucial impacts on activating the backdoor.
    \item We verify that attacks with static trigger pattern are transformation vulnerable, based on which we propose a simple, effective, and efficient transformation-based defense method.
    \item We propose an effective method to enhance the robustness against the change of trigger, and link the proposed enhancement with the physical attack.
\end{itemize}
}

\vspace{-0.6em}
\section{Related Work}
\vspace{-0.6em}

\subsection{Backdoor Attacks}
Backdoor attack is an emerging research area, which raises serious concerns about training with third-party datasets or platforms. Similar to the data poisoning \citep{biggio2012poisoning,alfeld2016data,liu2019unified}, backdoor adversary also tampers the training process to achieve their goals. However, these methods have different purposes. Specifically, the target of data poisoning is to degrade the model's performance on 
benign inputs, whereas the backdoor attack is aiming to misclassify inputs as a target class when the input is manipulated by adding a backdoor trigger. Meanwhile, the infected model can still correctly recognize the label for any benign sample. 

The backdoor attack was first proposed in \citep{gu2017badnets,gu2019badnets}. After that, \citep{chen2017targeted} first discussed the invisible backdoor attack, where the trigger is visually imperceptible. Recently, \citep{turner2019label} proposed a more stealthy attack approach, dubbed label-consistent attack, where the target label of poisoned samples is consistent with their ground-truth label. Several other backdoor attacks have also been proposed for different purposes \citep{liu2017trojaning,yao2019latent,bagdasaryan2018backdoor}. Except for image classification, backdoor attacks were also demonstrated to be effective towards other tasks \citep{zhao2020clean,kurita2020weight,zhai2020backdoor}. Although various backdoor attacks were proposed, most of them have a static trigger setting, and the research on their mechanisms and properties is left far behind.

\vspace{-0.3em}
\subsection{Backdoor Defenses}
\vspace{-0.3em}
To defend backdoor attacks, several empirical defense methods were proposed. These methods can be roughly divided into six main categories, including \emph{preprocessing-based defense} \citep{liu2017neural,gia2019februus,villarreal2020confoc}, \emph{model reconstruction based defense} \citep{liu2017neural,liu2018fine,zhao2020bridging}, \emph{trigger synthesis based defense} \citep{wangneural, qiao2019defending,zhu2020GangSweep}, \emph{model diagnosis based defense} \citep{kolouri2020universal,huang2020one,wang2020practical}, \emph{poison suppression based defense} \citep{hong2020effectiveness,du2019robust}, and \emph{sample filtering based defense} \citep{tran2018spectral,gao2019strip,javaheripi2020cleann}. Unfortunately, existing defenses either suffer from high complexity or relatively low clean accuracy. Not to mention that most of them have already been bypassed by subsequently adaptive attacks. How to better defend against backdoor attacks is still an important open question.


\vspace{-1em}
\section{The Property of Existing Attacks with Static Trigger}
\vspace{-0.3em}
\label{sec_limit}

\subsection{Backdoor Attack with Static Trigger}
We consider the scenario that the user cannot fully control the training process of the model $C(\cdot;w)$. Let $y_{target}$ denotes the target label, $\mathcal{D}_{train} = \{ (\bm{x}, y) \}$ with $\bm{x} \in \{0,1,\ldots, 255\}^{C\times W \times H}$ indicates the (benign) training set. 
The target of backdoor attack is to obtain an \emph{infected model}, which performs well on benign tesing images whereas it may have been injected some insidious backdoors. 

The typical process of the backdoor attack has two main steps: {\bf (1)} generate the poisoned image $\bm{x}_{poisoned}$ with target label $y_{target}$; {\bf (2)} Adopte both the benign and poisoned samples for training.

\noindent \textbf{The Generation of Poisoned Images.} 
As stated above, generating poisoned images is the first step of backdoor attack. Specifically, the poisoned image $\bm{x}_{poisoned}$ is generated through a \emph{stamping process} $S$ based on the trigger $\bm{x}_{trigger}$ and the benign image $\bm{x}$, $i.e.$,
\begin{equation}
    \bm{x}_{poisoned} = S(\bm{x};\bm{x}_{trigger}) = (\bm{1}-\bm{\alpha}) \otimes \bm{x} + \bm{\alpha} \otimes \bm{x}_{trigger},  
    \label{eq: poisoned image}
\end{equation}
where $\bm{\alpha} \in [0,1]^{C \times W \times H}$ is a trade-off hyper-parameter and $\otimes$ indicates the element-wise product.
%

\noindent \textbf{Training Process.} 
We denote  $\mathcal{D}_{benign}$ as all benign samples used for backdoor training ($\mathcal{D}_{benign} \subset \mathcal{D}_{train}$), and denote the set of poisoned samples as $\mathcal{D}_{poisoned}=\{(\bm{x}_{poisoned}, y_{target})\}$. Both of them are utilized to train the model, as follows
\begin{equation}\label{obj_attack}
    \min_{w} \mathbb{E}_{(x,y) \in \mathcal{D}_{poisoned} \cup \mathcal{D}_{benign}}  \mathcal{L}\left(C(\bm{x};w), y\right),
\end{equation}
where $\mathcal{L}(\cdot)$ indicates the loss function, such as the cross entropy. 

\vspace{-0.2em}
\subsection{The Effects of Different Characteristics}
\label{sec_char}
\begin{figure}[ht]
\begin{minipage}[b]{0.55\linewidth}
\begin{defn}[Minimum Covering Box]
The minimum covering box is defined as the minimum bounding box in the poisoned image covering the whole trigger pattern ($i.e.$, all non-zero $\bm{\alpha}$ entries).
\end{defn}

\vspace{0.4em}

\begin{defn}[Two Characteristics of Backdoor Trigger]
\label{def: three characteristics}

A trigger can be defined by two independent characteristics, including \textbf{location} and \textbf{appearance}. Specifically, \textbf{location} is defined by the position of the pixel at the bottom right corner of the minimum covering box, and \textbf{appearance} is indicated by the color value and the specific arrangement of pixels corresponding to non-zero $\bm{\alpha}$ entries in the minimum covering box.
\end{defn}
\end{minipage}
\hfill
\begin{minipage}[b]{0.43\linewidth}
 \centering
 \vspace{-0.5em}
 \includegraphics[width=\textwidth]{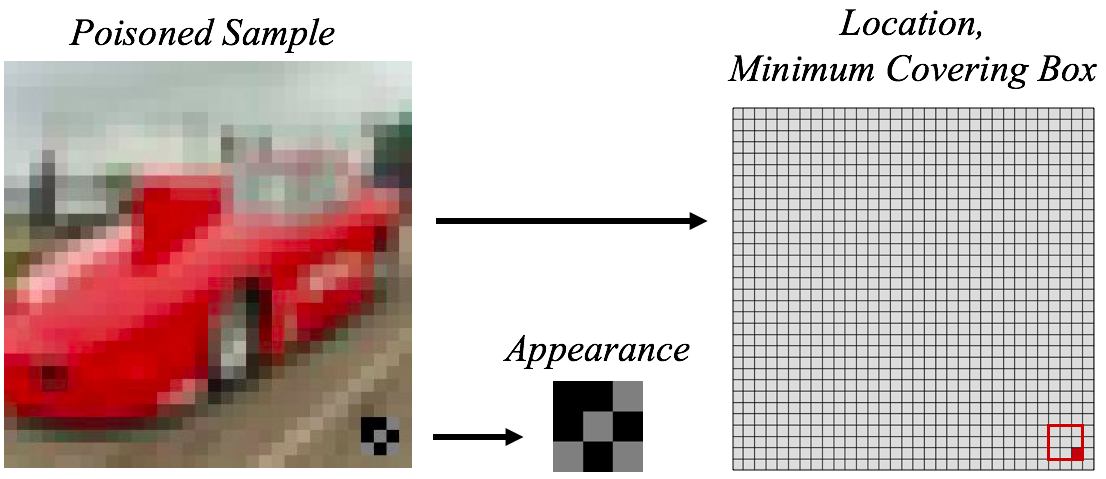}
 \vspace{-1.2em}
 \caption{The illustration of characteristics of the backdoor trigger. The red box represents the boundary of the minimum covering box, and the red pixel indicates the trigger location.} 
 \label{ill_TwoChars}
\end{minipage}
\vspace{-0.8em}
\end{figure}

One backdoor trigger can be specified by two independent characteristics, including \emph{location} and \emph{appearance}. 
To study their individual effects to backdoor attack, we firstly present their accurate definitions in Definition \ref{def: three characteristics}. One illustrative example is also shown in Figure \ref{ill_TwoChars}.

\begin{figure}[ht]
\vspace{-1.2em}
\begin{minipage}[b]{0.5\linewidth}
    \centering
    \subfigure[VGG-19]{
    \includegraphics[width=0.48\textwidth]{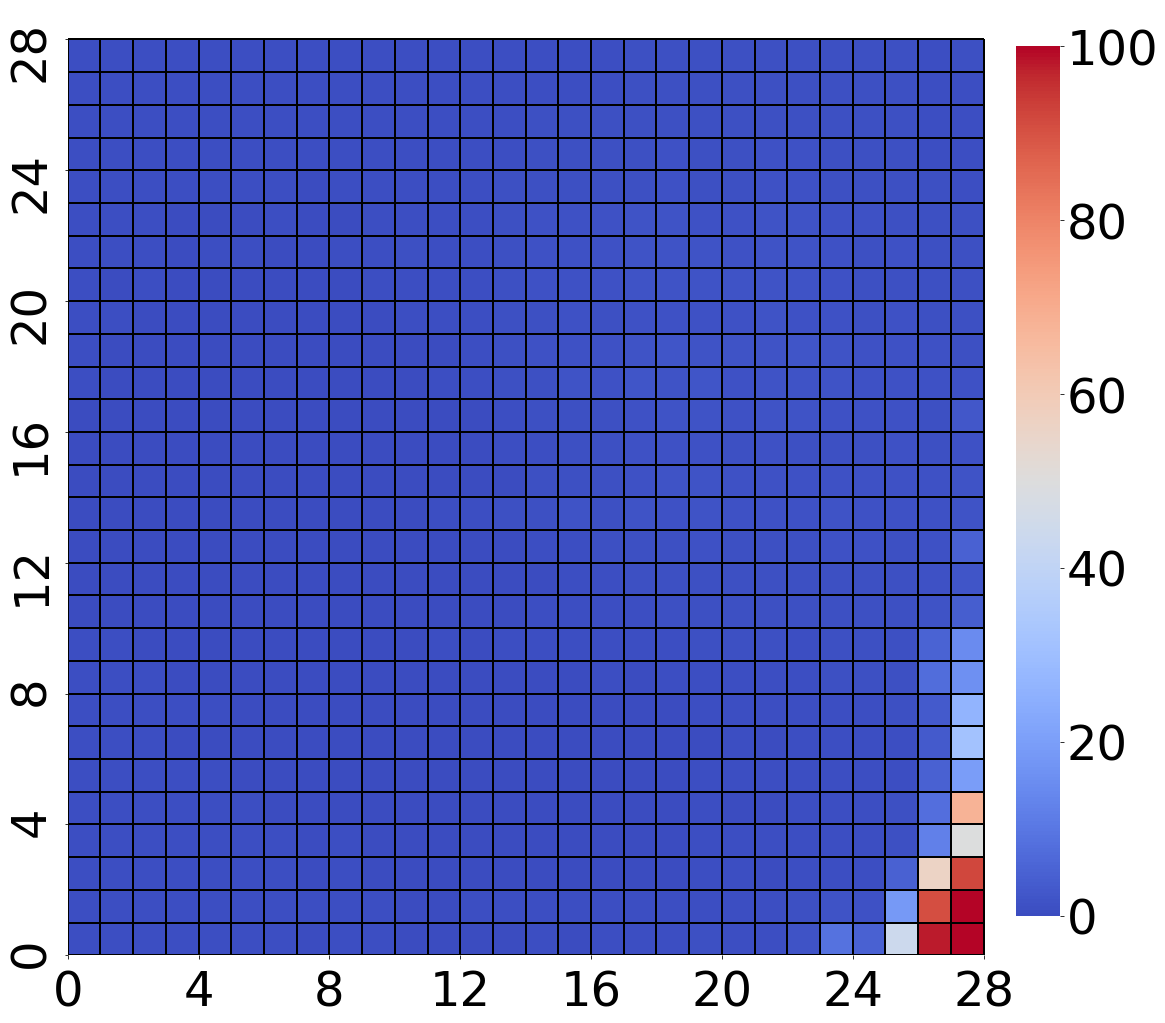}}
    \subfigure[ResNet-34]{
    \includegraphics[width=0.48\textwidth]{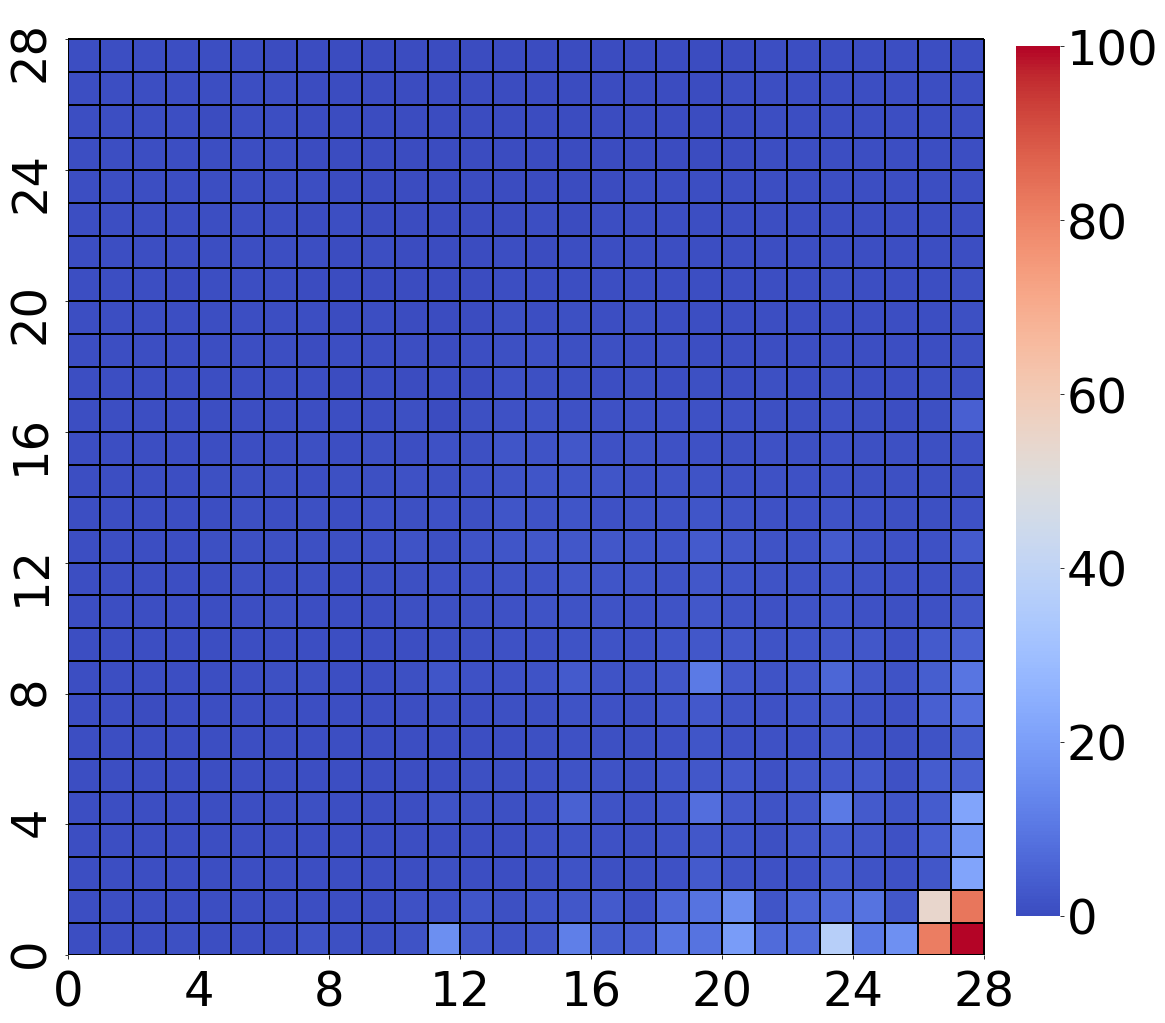}}
    \vspace{-0.6em}
    \caption{The heatmap of the attack success rate when the trigger is in different position at attacked images. The right corner is the position of the trigger in the poisoned images used for training.}
\label{fig_position}
\end{minipage}\quad
\begin{minipage}[b]{0.47\linewidth}
    \centering
    \includegraphics[width=0.9\textwidth]{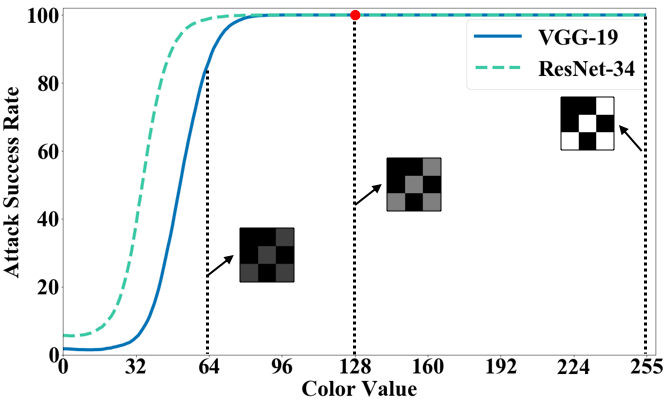}
    \vspace{-0.6em}
    \caption{ASR and appearance of the trigger with different non-zero color value in attacked images. The red dot indicates the ASR of trigger with original color value (128 pixel).}
    \label{fig_color}
\end{minipage}
\vspace{-1.2em}
\end{figure}

\noindent \textbf{Evaluation Criteria of Attacks.} 
We adopt the attack performance to measure the effect, which is specified as the \emph{attack success rate} (ASR). It is defined as the accuracy of attacked images predicted by the infected classifier $C(\cdot;\hat{w})$ with stamping process $S$, $i.e.$, 
\begin{equation}
   ASR_{C}(S) =  \Pr_{(\bm{x},y) \in \mathcal{D}_{test}}\left[C\left(S(\bm{x};\hat{w})\right)=y_{target}\left.\right| y \neq y_{target} \right].
\end{equation}
For the sake of brevity, we will use $ASR(\cdot)$ instead, if specifying $C(\cdot;\hat{w})$ is not necessary.

\noindent \textbf{Settings.} In the following experiments in this section, we use BadNets \citep{gu2019badnets} as an example to study the effects of location and appearance. We use VGG-19 \citep{simonyan2014very} and ResNet-34 \citep{he2016deep} as the model structure, and conduct experiments on CIFAR-10 dataset \citep{krizhevsky2009learning}. The trigger is a $3\times3$ black-gray square, as shown in Figure \ref{ill_TwoChars}. More details are shown in Appendix \ref{app:twochar}.


\noindent \textbf{The Effect of Location.}
While preserving the appearance of the trigger, we change its location in inference process to study its effect to the attack performance. As shown in Figure \ref{fig_position}, when moving the location with a small distance ($2 \sim 3$ pixels, less than $10\%$ of the image size), the ASR will drop sharply from $100\%$ to below $50\%$. It tells that the attack performance is sensitive to the location of the backdoor trigger on the attacked image in the inference process.

\noindent \textbf{The Effect of Appearance.} 
While keeping the location of the trigger, we change its appearance in the inference stage to study the appearance's effect on the attack performance. The appearance could be modified by changing the shape or the pixel values of the trigger. For the sake of simplicity, here we only consider the change of pixel values. 
Specifically, there are only two values of the pixels within the trigger, $i.e.$, 0 and 128. 
We change the value 128 to different values from 0 to 255. The ASR scores corresponding to different pixel values are plotted in Figure \ref{fig_color}. As shown in the figure, the ASR degrades sharply along with the decreasing of non-zero pixel values, while is not influenced when the non-zero pixel values are increased. 
According to this simple experiment, it is difficult to describe the exact relationship between the change of appearance and the attack performance, since the change modes of appearance are rather diverse. However, it at least tells that the attack is sensitive to the difference of appearance between the trigger on the attacked testing image and that used in training. More explorations about this phenomenon will be discussed in our future work.

\vspace{-0.7em}
\section{Further Explorations of the Property}
\vspace{-0.6em}

The studies presented in Section \ref{sec_limit} demonstrate that the backdoor attack is sensitive to the difference between the training trigger and the testing trigger. 
It gives us two further questions: {\bf (1)} Is it possible to utilize such a sensitivity to defend the current backdoor attacks with static trigger? {\bf (2)} How to enhance the robustness of the backdoor attack to the change of trigger? We propose two simple yet effective approaches to answer this two questions in Section \ref{sec_transdefense} and Section \ref{sec_enhancement}, respectively.

\vspace{-0.4em}
\subsection{Backdoor Defense via Transformations}
\vspace{-0.3em}
\label{sec_transdefense}

The answer to the first question is to change the location or appearance of the trigger in the inference process, such that the modified trigger may fail to activate the backdoor hidden in the model. However, since the user doesn't have the information about the trigger, it is impossible to exactly manipulate the trigger. Instead, we propose a transformation-based defense by changing the whole image with some transformations (\eg, flipping or scaling), as shown in Definition \ref{def_defense}.

\begin{figure*}[ht]
 \centering
 \vspace{-1em}
 \includegraphics[width=0.85\textwidth]{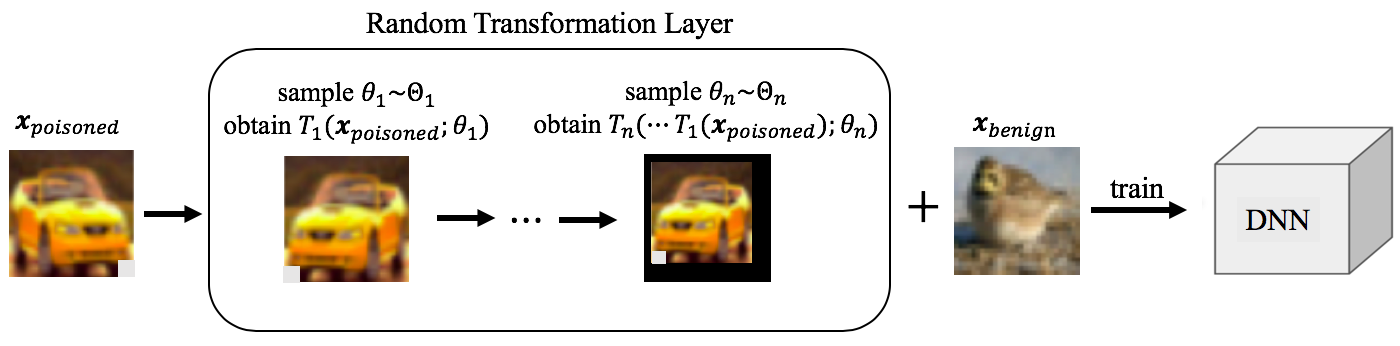}
  \vspace{-0.8em}
 \caption{The pipeline of the proposed transformation-based attack enhancement.} 
 \label{ill_exp}
 \vspace{-1.2em}
\end{figure*}

\begin{defn}[transformation-based defense]\label{def_defense}
    The transformation-based defense is defined as introducing a transformation-based pre-processing module on the testing image before prediction, $i.e.$, instead of predicting $\bm{x}$, it predicts $T(\bm{x})$, where $T(\cdot)$ is a transformation. 
\end{defn}

This simple defense method enjoys several advantages: {\bf (1)} it is efficient since it only requires the transform the testing image; {\bf (2)} it is attack-agnostic, therefore it can defend different attacks with static trigger simultaneously; {\bf (3)} it is data-free and model-free, $i.e.,$ compared with most existing defense methods, the defender does not need to have any clean samples or modify the model parameters. Accordingly, it would be the primary defense approach when adopting the third-party API of DNNs-based algorithms. These advantages will be further verified in Section \ref{sec_exp}.

In particular, we suggest to use spatial transformations for the defense, since it may probably change the location and appearance of the trigger simultaneously, while the location has a direct connection to the backdoor activation. A more comprehensive discussion about the defense with different transformations will be demonstrated in Appendix \ref{defense:non-spatial}.

\vspace{-0.15em}
\subsection{Transformation-based Enhancement and Physical Backdoor Attack}
\vspace{-0.15em}
\label{sec_enhancement}

In this section, we discuss how to enhance the transformation-robustness of existing attacks, and its connection with the physical attack. 

\begin{defn} [Transformation Robustness]
The transformation robustness of attack with stamping process $S$ under transformation $T(\cdot;\bm{\theta})$ (with parameter $\bm{\theta}$), the $R_T(S)$, is defined as the attack success rate after the transformation $T$, $i.e.$,
\begin{equation}
    R_T(S) = ASR(T(S)),
\end{equation}
\end{defn}
where 
$$
ASR(T(S))=\Pr_{(\bm{x},y) \in \mathcal{D}}\left[C\left(T(S(\bm{x}))\right)=y_{target}\left.\right| y \neq y_{target} \right].
$$
Note that $R_T(S)\in [0,1]$. The larger value of $R_T(S)$ indicates the higher robustness towards transformation $T$. 
Besides, the transformation could also be a {\it compound transformation} $T(\cdot;\bm{\theta})$ of a sequence of basic transformation $\{T(\cdot;\theta_i)\}_{i=1}^n$, $i.e.,$ $T(\cdot;\bm{\theta}) = T_n(T_{n-1}(\cdots T_1(\cdot;\theta_1);\theta_{n-1});\theta_n)$.

\comment{
\begin{defn} [Compound Transformation]
The compound transformation $T(\cdot;\bm{\theta})$ 
of a sequence of transformation with parameter $\theta_i$, the $\{T(\cdot;\theta_i)\}_{i=1}^n$, 
is formulated as the composition of a sequence of transformation functions, $i.e.$,
\begin{equation}
    T(\cdot;\bm{\theta}) = T_n(T_{n-1}(\cdots T_1(\cdot;\theta_1);\theta_{n-1});\theta_n),
\end{equation}
where $\bm{\theta}=(\theta_1, \cdots,\theta_n)$.
\label{def: compound transform}
\end{defn}
}

\comment{
Once the transformation is known by the attacker, a simple method can be used to enhance the attack robustness. 
Specifically, the generation of poisoned images, which was defined in Eq. (\ref{eq: poisoned image}), is updated to as follows:
\begin{equation}
  \hspace{-0.3em} \bm{x}_{poisoned}' = T(\bm{x}_{poisoned};\bm{\theta}) = T(S(\bm{x};\bm{x}_{trigger});\bm{\theta}),  
\end{equation}
which means that the poisoned images are pre-processed through the compound transformation, before being fed into the training process.
Accordingly, similar to Eq. (\ref{obj_attack}), the training objective is updated as follows:
\begin{equation}\label{obj_attack_enhance}
    \min_{w} \mathbb{E}_{(\bm{x},y) \in \mathcal{D}_{poisoned}^{(T)} \cup \mathcal{D}_{benign}}  \mathcal{L}\left(C(\bm{x};w), y\right),
\end{equation}
where $\mathcal{D}_{poisoned}^{(T)} = \{(\bm{x}_{poisoned}', y_{target})\}$.
}

The key issues for improving transformation-robustness are how to determine the compound transformation and the corresponding parameter $\boldsymbol{\theta}$ used by defenders. 
In practice, the attacker is difficult to know the exact transformations. Even the adopted transformations are revealed to the attacker, the exact parameters in transformations cannot be known, as there may be randomness in practice (\ie, different scaling factors in scaling transformation). 
To tackle this difficulty and to ensure the attack capability towards different possible transformation-based defenses, we specify $T$ with the set of some common transformations. 
For each $T_i$, if there may be randomness in practice, then we define a value domain $\Theta_i$ for $\theta_i$. $\Theta_i$ is parameterized by the maximal transformation size $\epsilon$, 
$i.e.$,
$$
    \Theta_i = \{\theta|dist_i(\theta, I) \leq \epsilon_i\}, 
$$
where $dist_i(\cdot, \cdot)$ is a given distance metric for $T_i$, and $I$ indicates the identity transformation. 

Consequently, the compound transformation used in the enhanced attack is specified as $\mathcal{T}= \{T(\cdot;\bm{\theta})|\bm{\theta} \in \prod_{i=1}^{n}\Theta_i\}$. 
Then, the training objective of the enhanced attack is formulated as 
\begin{equation}\label{obj_attack_class_enhance}
    \min_{w} \mathbb{E}_{\bm{\theta}} \left[\mathbb{E}_{(\bm{x},y) \in \mathcal{D}_{poisoned}^{(T(\cdot;\bm{\theta}))} \cup \mathcal{D}_{benign}} \left[\mathcal{L}\left(C(\bm{x};w), y\right)\right]\right].
\end{equation}

\begin{table*}[ht]
\center
\scriptsize
\vspace{-1.5em}
\caption{Comparison of different backdoor defenses on CIFAR-10 dataset. `Clean' and `ASR' indicates the accuracy (\%) and attack success rate (\%) on testing set, respectively. The boldface indicates the best results among all preprocessing based defenses.}
\label{defense}
\scalebox{0.93}{
\begin{tabular}{c|cc|cc|cc|cccccc}
\hline
\multicolumn{1}{c|}{Model Architectures $\rightarrow$} & \multicolumn{6}{c|}{VGG-19}                                                                                                                                                                       & \multicolumn{6}{c}{ResNet-34}                                                                                                                                                                                                              \\ \hline
\multicolumn{1}{c|}{Attack Methods $\rightarrow$} & \multicolumn{2}{c|}{BadNets}                                    & \multicolumn{2}{c|}{Blended Attack}                            & \multicolumn{2}{c|}{Consistent Attack}                         & \multicolumn{2}{c|}{BadNets}                                                        & \multicolumn{2}{c|}{Blended Attack}                                                 & \multicolumn{2}{c}{Consistent Attack}                          \\ \cline{2-13} 
\multicolumn{1}{c|}{Defense Methods $\downarrow$} & Clean & \begin{tabular}[c]{@{}c@{}}ASR\end{tabular} & Clean & \begin{tabular}[c]{@{}c@{}}ASR\end{tabular} & Clean & \begin{tabular}[c]{@{}c@{}}ASR\end{tabular} & Clean & \multicolumn{1}{c|}{\begin{tabular}[c]{@{}c@{}}ASR\end{tabular}} & Clean & \multicolumn{1}{c|}{\begin{tabular}[c]{@{}c@{}} ASR \end{tabular}} & Clean & \begin{tabular}[c]{@{}c@{}} ASR \end{tabular} \\ \hline
Standard          & 91.9  & 100                                                     & 91.5     & 100                                                   & 91.3  & 95.6                                                   & 94.1  & \multicolumn{1}{c|}{100}                                                   & 93.1     & \multicolumn{1}{c|}{100}                                                   & 93.1  & 98.7                                                   \\ \hline

Fine-Pruning          & 91.3  & 0.7                                                     & 83.6     & 0.2                                                   & 72.6  & 0.1                                                   & 92.1  & \multicolumn{1}{c|}{0}                                                   & 91.9     & \multicolumn{1}{c|}{0.3}                                                   & 92.0  & 18.9                                                   \\
Neural Cleanse        & 83.3     & 0.6                                                       & 90.6     & 0.4                                                      & 86.4     & 0.7                                                      & 91.4     & \multicolumn{1}{c|}{0.7}                                                      & 91.4     & \multicolumn{1}{c|}{0.5}                                                      & 91.2     & 1.4                                                      \\ \hline
Auto-Encoder           & 86.4  & 2.1                                                     & 86.0     & 1.7                                                      & 85.4  & \textbf{2.3}                                                    & 87.5  & \multicolumn{1}{c|}{2.7}                                                    & 87.2     & \multicolumn{1}{c|}{1.9}                                                      & 88.4  & \textbf{2.1}                                                    \\ 
Flip (Ours)                 & \textbf{91.0}    & \textbf{1.1}                                                     & \textbf{91.1}     & \textbf{0.9}                                                      & \textbf{90.5}  & 95.7                                                   & \textbf{93.6}  & \multicolumn{1}{c|}{\textbf{0.8}}                                                    & \textbf{92.8}     & \multicolumn{1}{c|}{\textbf{0.8}}                                                      & \textbf{92.3}  & 98.8                                                   \\
ShrinkPad-4 (Ours)         & 87.6  & 1.6                                                    & 88.3     & 1.8                                                      & 87.5  & 3.7                                                    & 91.4  & \multicolumn{1}{c|}{1.5}                                                   & 90.6     & \multicolumn{1}{c|}{1.8}                                                      & 89.9  & 4.8                                                   \\ \hline
\end{tabular}
}
\vspace{-1.5em}
\end{table*}

To solve the problem (\ref{obj_attack_class_enhance}) exactly, attackers need to conduct the training process with all possible transformed variants, which is computation-consuming. Instead, 
we propose a sampling-based method for efficiency. Specifically, for each poisoned image, to handle the expectation over all possible configurations of  $\bm{\theta}$, we sample one configuration, \ie, $\bm{\theta} \sim \prod_{i=1}^{n}\Theta_i$, based on which we transform the original images. Then, we use the transformed poisoned images and benign images for training. The training process of the proposed enhanced attack is briefly illustrated in Figure \ref{ill_exp}.

\textbf{The connection between the proposed attack enhancement and physical attack. } In real-world scenarios, the testing image may be acquired by some digitizing devices (\eg, camera), rather than be directly provided in the digital space.  
In those scenarios, the trigger should be stamped on the object, which is then digitized by the camera to fool the model. It is dubbed {\it physical attack}. Some recent works \citep{schwarzschild2020just, wenger2020backdoor} demonstrated that existing backdoor attacks are vulnerable under the scenario of physical attack. It is due to that the relative distance and angle between the photo and the camera is varied in practice, therefore the location and appearance of the trigger in the digitized attacked image may be different from that of the trigger used for training. These spatial variations in physical scenarios can be approximated
by some widely used transformations (\eg, spatial transformations), which have been incorporated into proposed transformation-based enhancement. Thus, it is expected that the proposed transformation-based enhancement can serve as an effective physical attack, which will be futher verified in Section \ref{sec_physical}.

\vspace{-0.3em}
\section{Experiment}
\vspace{-0.4em}
\label{sec_exp}

\subsection{Transformation-based Defense}
\label{transdefense}
\vspace{-0.2em}

In this section, we verify the effectiveness of the proposed defense with spatial transformations. We examine two simple spatial transformations, including left-right flipping (dubbed {\it Flip}), and padding after shrinking (dubbed {\it ShrinkPad}). Specifically, ShrinkPad consists of shrinking (based on bilinear interpolation) with a few pixels ($i.e.$, shrinking size), and random zero-padding around the shrunk image. The results of defense with non-spatial transformations will be shown in Appendix \ref{app:sec_defense}.

\textbf{Settings. } We use three representative backdoor attacks, including BadNets \citep{gu2017badnets}, attack with blended strategy \citep{chen2017targeted} (dubbed Blended Attack), and label consistent backdoor attack \citep{turner2019label} (dubbed Consistent Attack) to evaluate the performance of backdoor defenses. For BadNets and Blended Attack, the trigger is a $3\times3$ black-white square, which is similar to the one used in Section \ref{sec_char}.
For defense comparison, we select four important baseline, including fine-pruning \citep{liu2018fine}, neural cleanse \citep{wangneural}, auto-encoder based defense (dubbed Auto-Encoder) \citep{liu2017neural}, and standard training (dubbed Standard).

\textbf{Results. } As shown in Table \ref{defense}, the proposed defense is effective. Specifically, ShrinkPad with 4 pixels shrinking size could decrease the ASR by more than $90\%$ in all cases. Flip also shows satisfied defense performance towards BadNets and Blended attacks. But it doesn't work on defending against Consistent Attack, due to the symmetrical trigger used in Consistent Attack. Compared with the state-of-the-art preprocessing based method ($i.e.,$ Auto-Encoder), the proposed method has higher clean accuracy and lower ASR in general. Besides, its performance is even on par with Fine-Pruning and Neural Cleanse, which require stronger defensive capabilities (\ie, modify the model parameters and access to benign samples). Moreover, the proposed method is more efficient compared with other baseline methods, and is even more effective when the backdoor trigger is the universal adversarial perturbation \citep{moosavi2017universal}. It will be shown in Appendix \ref{app:moredefense}.


\vspace{-0.4em}
\subsection{Attack Enhancement} \label{sec:attack_enhance}
\vspace{-0.3em}
\noindent \textbf{Settings. } 
In the enhanced backdoor attack, we adopt random Flip followed by random ShrinkPad in the random transformation layer. There is only one hyper-parameter in the enhanced attack, $i.e.$, the maximal shrinking size, which is set to 4 pixels in our experiments. Other settings are the same as those used in Section \ref{transdefense}. More setting details and an ablation study about the effect of the hyper-parameter are shown in Appendix \ref{app:setattack} and Appendix \ref{app:aba}, respectively.

\begin{table*}[t]
\center
\vspace{-0.8em}
\scriptsize
\caption{The comparison between standard backdoor attacks and enhanced backdoor attacks from the aspect of attack success rate against different transformation-based defenses. }
\begin{tabular}{c|cccc|cccc}
\hline
Model Architectures $\rightarrow$                 & \multicolumn{4}{c|}{VGG-19}                                                       & \multicolumn{4}{c}{ResNet-34}                                                    \\ \hline
                        Attacks $\downarrow$, Defenses $\rightarrow$   & Standard     & Flip           & ShrinkPad-2  & ShrinkPad-4         & Standard     & Flip           & ShrinkPad-2  & ShrinkPad-4  \\ \hline
BadNets                       & \textbf{100.0} & 1.1            & 22.7           & 1.6                      & \textbf{100.0} & 0.8            & 14.9            & 1.5           \\
BadNets+             & \textbf{100.0} & \textbf{100.0} & \textbf{100.0} & \textbf{100.0}  & \textbf{100.0} & \textbf{100.0} & \textbf{100.0} & \textbf{100.0} \\ \hline
Blended Attack               & \textbf{100.0} & 0.9            & 40.8            & 1.8                      & \textbf{100.0} & 0.8            & 18.2            & 1.8            \\
Blended Attack+     & 99.9           & \textbf{99.9}  & \textbf{100.0} & \textbf{98.7}   & \textbf{100.0} & \textbf{100.0} & \textbf{100.0} & \textbf{99.5}  \\ \hline
Consistent Attack            & \textbf{95.6}           & \textbf{95.7}           & 67.1           & 3.7                      & \textbf{98.7}           & \textbf{98.8}          & 24.2           & 4.8            \\
Consistent Attack+    & 86.0     & 86.3     & \textbf{97.2}     & \textbf{90.9}     & 96.4     & 97.3     & \textbf{97.4}     & \textbf{98.7}     \\ \hline
\end{tabular}
\label{tab_EnhancedAttack}
\vspace{-1em}
\end{table*}

\begin{figure*}[ht]
\centering
\vspace{-0.4em}
\subfigure[Standard Backdoor Attack]{
\includegraphics[width=0.49\textwidth]{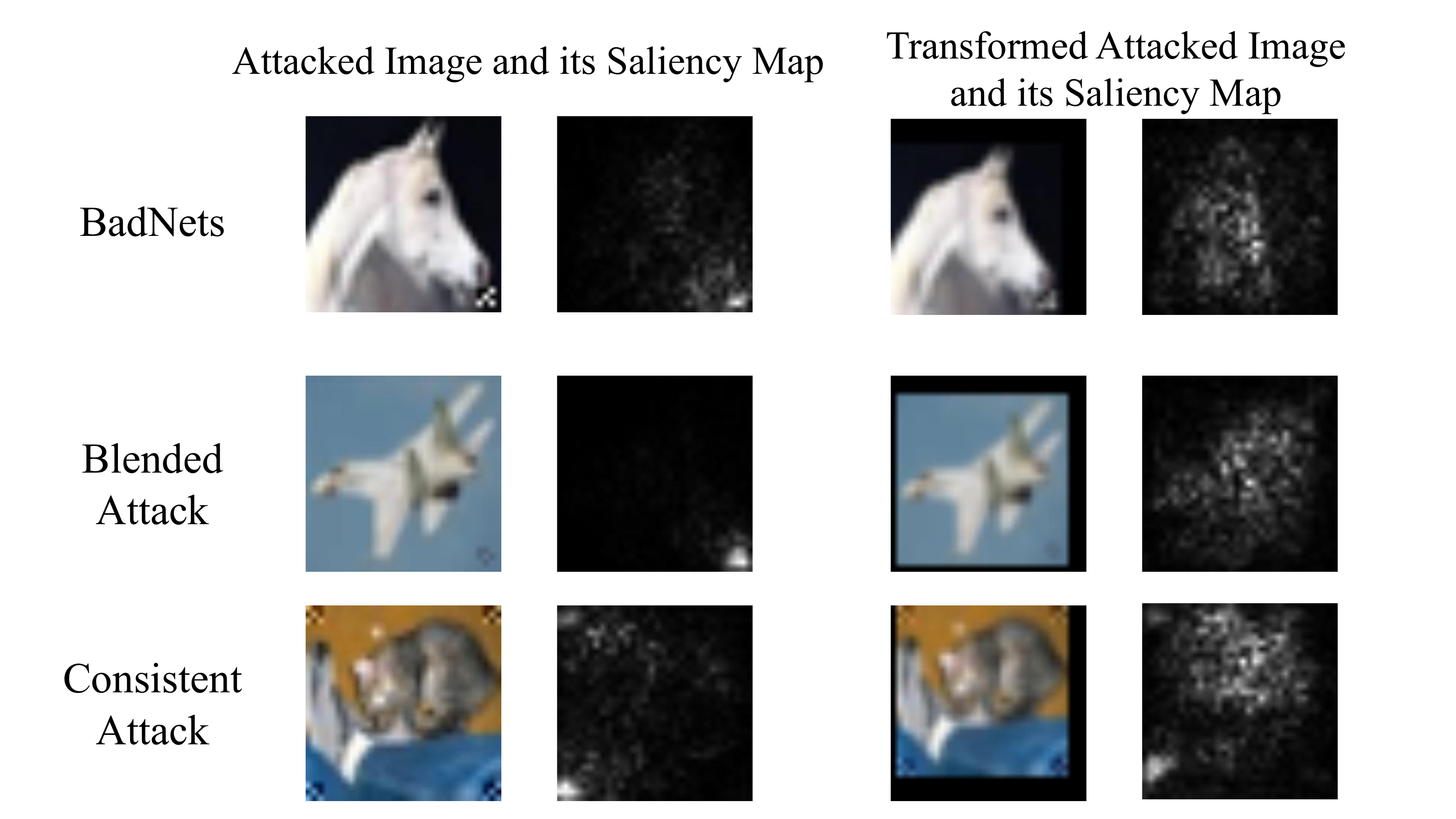}}
\subfigure[Enhanced Backdoor Attack]{
\includegraphics[width=0.49\textwidth]{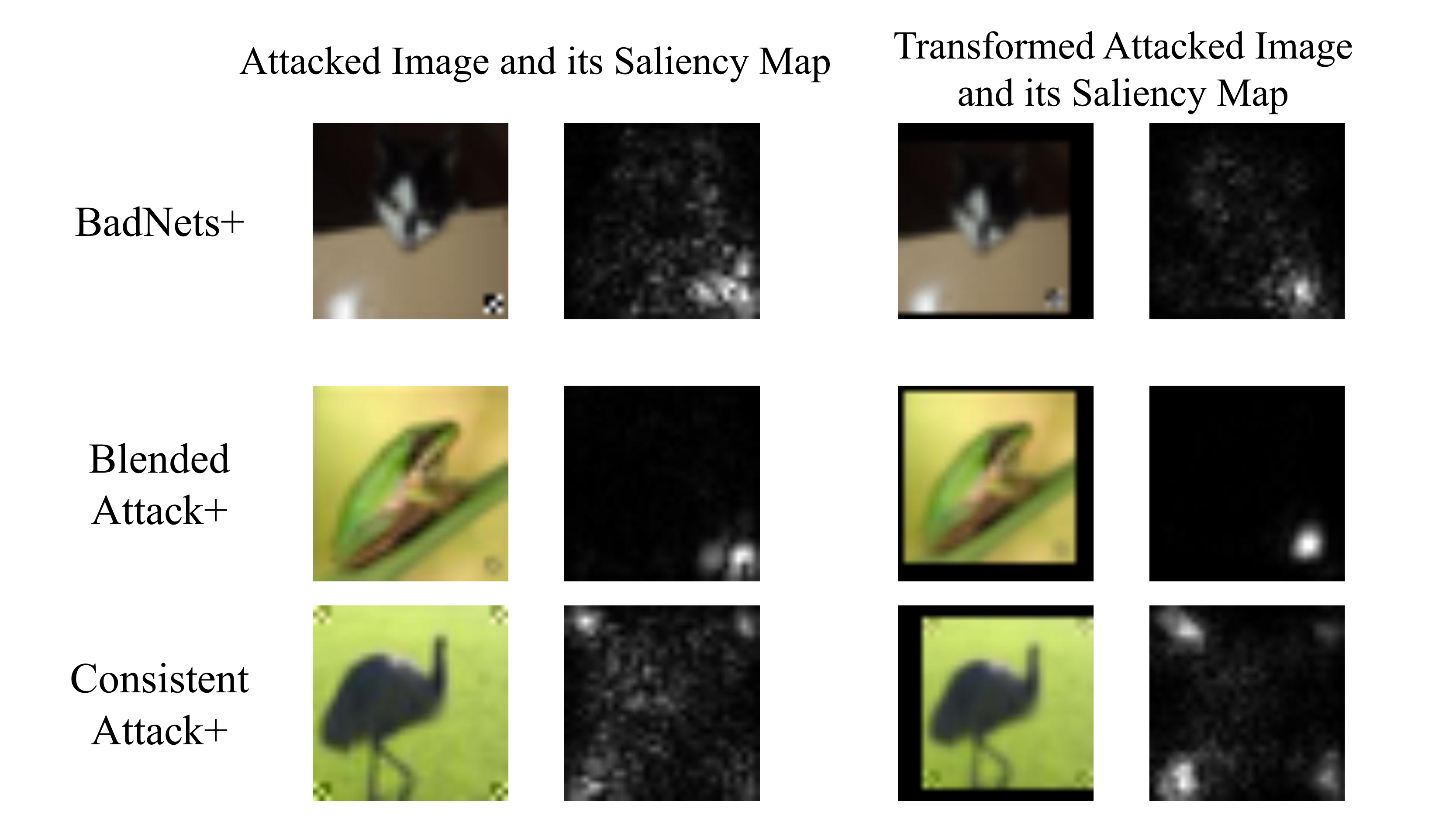}}
\vspace{-0.8em}
\caption{The saliency maps of images under standard and enhanced backdoor attacks.} 
\label{fig_saliency}
\vspace{-1em}
\end{figure*}

\noindent \textbf{Results. } 
As demonstrated in Table \ref{tab_EnhancedAttack}, enhanced backdoor attacks can still achieve a high ASR even under the defenses with spatial transformations. Specifically, the ASR of enhanced backdoor attacks is better than the one of their corresponding standard attack under defenses in almost all cases. Especially under ShrinkPad with shrinking 4 pixels, the ASR improvement of enhanced attacks is more than $85\%$ (mostly over $95\%$). The only exception is the Consistent Attack+ under Flip defense. It is partially due to the fact the trigger of Consistent Attack is symmetrical, as mentioned in Section \ref{transdefense}. Besides, 
compared to BadNets+ and Blended Attack+, Consistent Attack+ poisoned fewer images (see the attack settings), which is not favorable to the random trigger.

\vspace{-0.3em}
\subsection{The Differences between Backdoor Attack and Enhanced Attack}
\vspace{-0.3em}

In this section, we further explore the intrinsic difference between the standard backdoor attack and the enhanced backdoor attack (\ie, the attack with the enhancement). Specifically, we adopt the \emph{saliance map} \citep{simonyan2013deep} to understand their overall behaviors by identifying critical pixels of different images. Besides, we adopt the \emph{critical data routing paths} (CDRPs) \citep{wang2018cdrp} between different samples to discuss the layer-wise behaviors of different attacks. CDRPs are the paths contribute most to the prediction. More setting details are shown in Appendix \ref{app:moredetails}.

\begin{figure*}[t]
\centering
\subfigure[Standard Backdoor Attack]{
\includegraphics[width=0.483\textwidth]{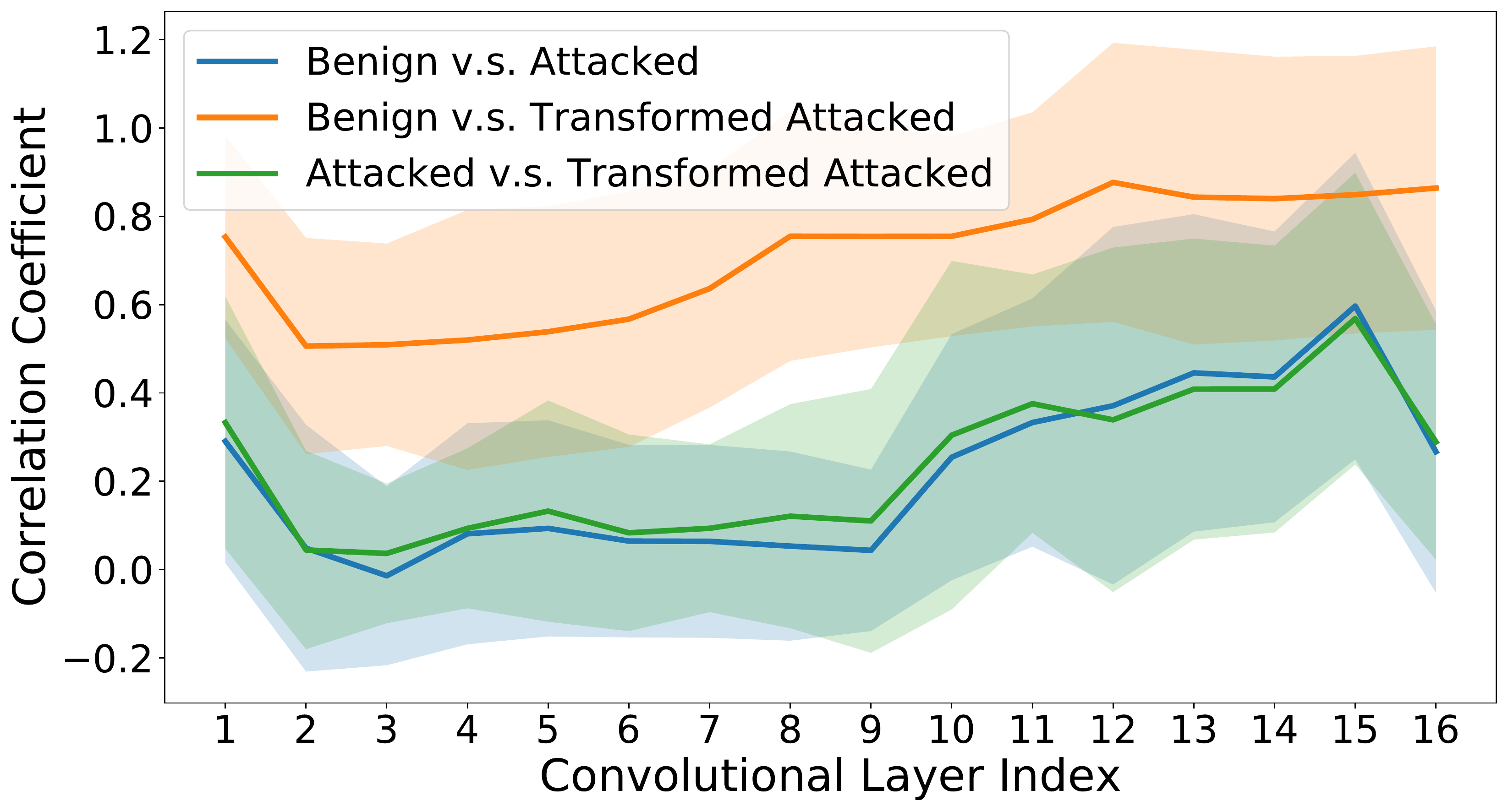}}
\subfigure[Enhanced Backdoor Attack]{
\includegraphics[width=0.483\textwidth]{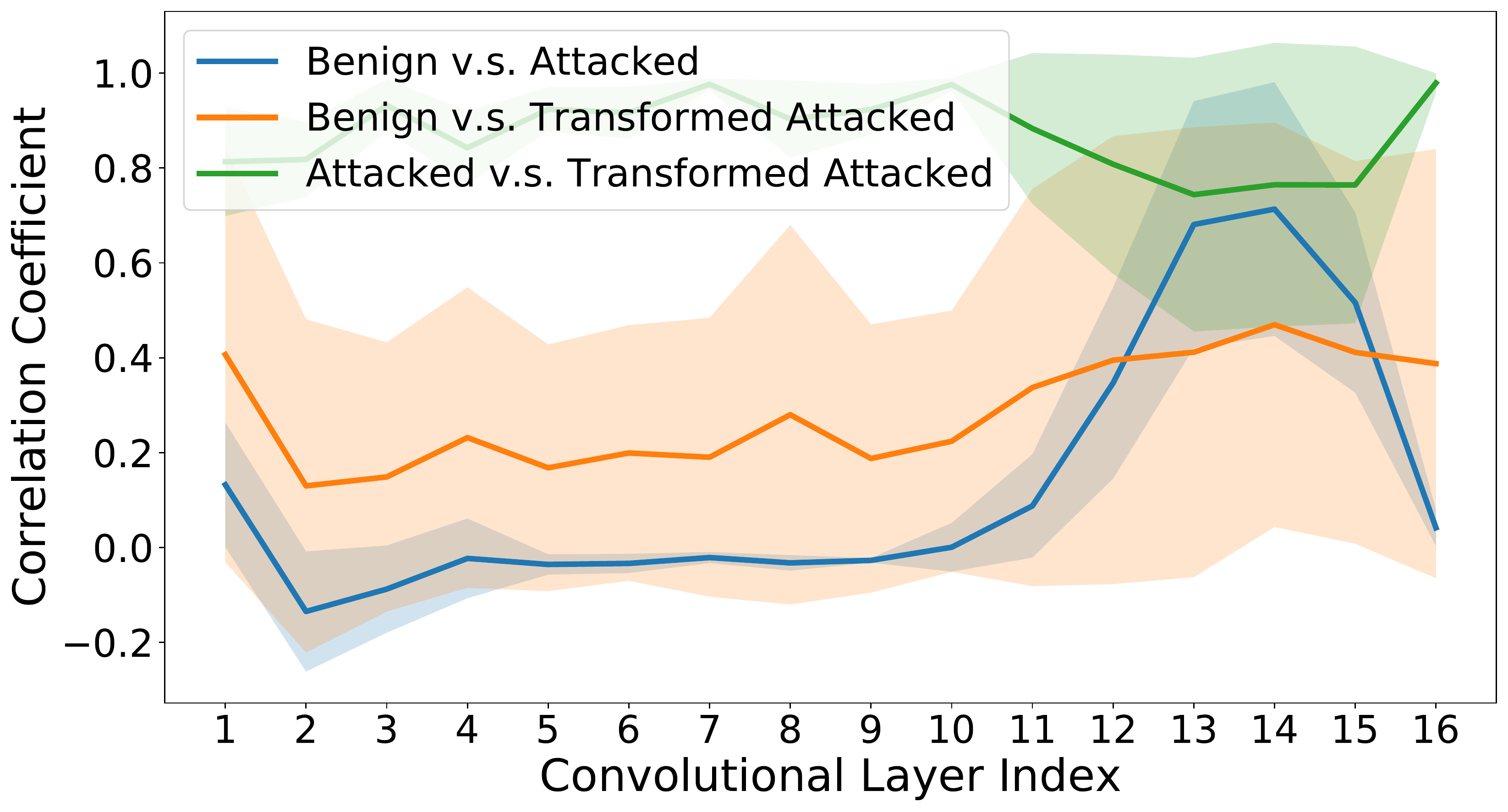}}
\caption{Layerwise correlation coefficients of critical data routing paths between (benign sample, attacked sample), (benign sample, transformed attacked sample), and (attacked sample, transformed attacked sample). The background color indicates the standard deviation over 100 samples.}
\label{Fig_CDP}
\vspace{-0.3em}
\end{figure*}

\begin{figure}[ht]
\begin{minipage}[h]{0.5\linewidth}
 \centering
 \includegraphics[width=\textwidth]{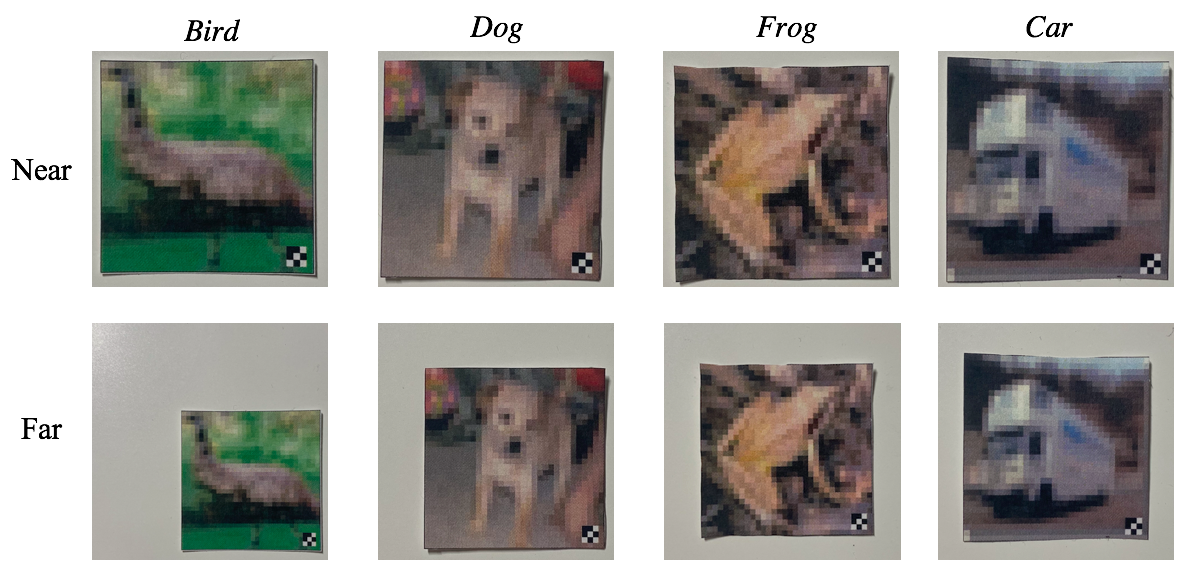}
 \vspace{-0.6em}
 \caption{The pictures of some printed CIFAR-10 images taken by a camera with different distances. All pictures are classified as `Deer' by the enhanced BadNets, whereas they will be classified as their benign label by the standard BadNets. }
 \label{phy1}
\end{minipage}\quad
\begin{minipage}[h]{0.47\linewidth}
 \centering
 \includegraphics[width=\textwidth]{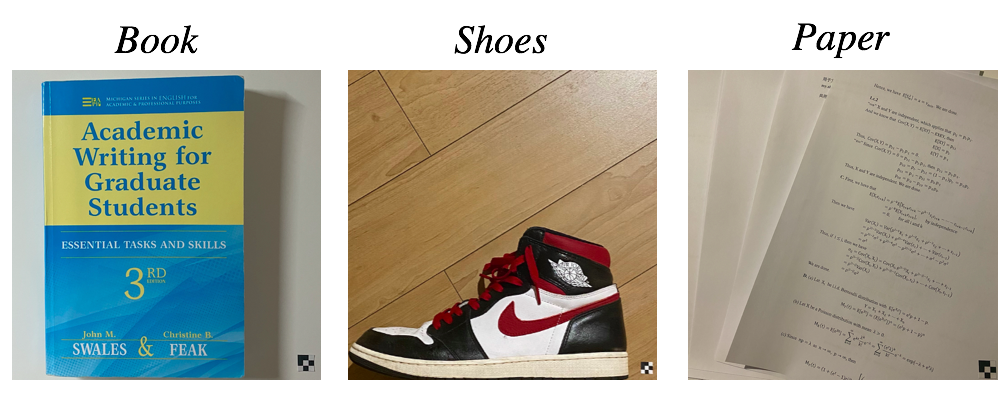}
 \caption{The picture of some out-of-sample images with the backdoor trigger taken by a camera. All pictures are classified as the target label `Deer' by the enhanced BadNets, whereas they will be classified as their benign label by the standard BadNets.}
 \label{phy2}
\end{minipage}
\vspace{-1em}
\end{figure}

As shown in Figure \ref{fig_saliency}, the saliency area of regularly ($i.e.$, non-transformed) attacked images mainly lies in the area of the backdoor trigger on both standard attacks and enhanced attacks, while the outline area of the object is not significantly activated. This phenomenon explains why these attacked samples can successfully mislead infected networks. 
Moreover, the saliency maps of transformed attacked images and those of regularly attacked images have significantly different patterns under standard attacks. For example, 
the saliency map of transformed attacked images mainly activates at object structure rather than at the backdoor trigger. In contrast, the saliency maps of attacked and transformed attacked images share certain similarities under enhanced attacks. The saliency maps of both types of attacked images concentrate on the area of the backdoor trigger. Those visualization results somewhat explain the different behaviors of standard attacks and enhanced attacks.

Figure \ref{Fig_CDP} (a) shows that the CDRPs of transformed attacked samples are similar with those of benign samples under standard backdoor attacks, corresponding to high correlation coefficients as demonstrated by the orange curve. In contrast, the CDRPs of another two pairs are different, corresponding to lower correlation coefficients shown in the blue and green curves.
This phenomenon explains why only attacked samples 
are classified as the target label, while both benign and transformed attacked samples are still classified as their ground-truth labels. 
Figure \ref{Fig_CDP} (b) shows that, under the enhanced backdoor attack, the CDRPs of attacked and transformed attacked samples are similar 
while their CDRPs are different from those of benign examples. 
This phenomenon is consistent with the result that both attacked and transformed attacked samples will be predicted by the enhanced attack as the target labels, which is different from their ground-truth labels. 


\vspace{-0.4em}
\subsection{Physical Backdoor Attack}
\vspace{-0.3em}
\label{sec_physical}

In this section, we further verify the effectiveness of the proposed attack enhancement under settings of the physical attack. Specifically, we evaluate BadNets and BadNets+ on the CIFAR-10 dataset. We randomly pick some testing images with backdoor trigger to take picture with differently relative location (near and far), as shown in Figure \ref{phy1}. Besides, we also take some \emph{out-of-sample} pictures that are totally different from the training images on CIFAR-10 dataset, as shown in Figure \ref{phy2}.

In the results of all figures, BadNets+ successfully enforces the prediction to the target label, while BadNets fails. 
Besides, the enhanced backdoor attack method is not only robust in the physical scenarios, but also generalizes well on out-of-sample images. This out-of-sample generalization is probably due to the strong relationship between the backdoor trigger and target label learned in the infected model, so that the impact of the non-trigger part is somewhat ignored by the model. 

\vspace{-0.5em}
\section{Conclusion}
\vspace{-0.4em}
In this paper, we explore the property of backdoor attacks.
%
%
We demonstrate that existing attacks with static trigger are transformation vulnerable, inspired by which we propose a simple yet effective transformation-based defense.
Besides, to reduce the transformation vulnerability of existing attacks, we propose a transformation-based enhancement by conducting the random spatial transformation on poisoned images before feeding into the training process. 
We also link the proposed attack enhancement to the physical attack and explore intrinsic differences between backdoor attack and enhanced attack. 
This work has shown that it is possible to develop simple yet effective defenses and attacks by utilizing some intrinsic properties. We hope that our approach could inspire more explorations on backdoor characteristics to help the design of more advanced methods.

\newpage

\comment{
\subsubsection*{Acknowledgments}
Use unnumbered third level headings for the acknowledgments. All
acknowledgments, including those to funding agencies, go at the end of the paper.
}

\bibliography{iclr2021_conference}
\bibliographystyle{iclr2021_conference}

\newpage

\appendix

\section{Settings for the Effects of Different Characteristics}
\label{app:twochar}

In this section, we illustrate the detailed settings in Section \ref{sec_char} of the main manuscript.

\noindent \textbf {Attack Setup.} We discuss the effects of trigger characteristics based on BadNets \citep{gu2019badnets} in these experiments. The trigger is a $3\times3$ black-gray square, as shown in Figure \ref{ill_TwoChars}. The trade-off hyper-parameter $\bm{\alpha}$ is set as $\bm{\alpha} \in \{0,1\}^{3\times32\times32}$. The values of $\bm{\alpha}$ entries corresponding to the pixels located in the minimum covering box are 1, while other values are 0.

\noindent \textbf {Training Setup.} We evaluate the effect with two popular CNN models, including VGG-19 \citep{simonyan2014very} and ResNet-34 \citep{he2016deep}, on the benchmark database CIFAR-10 \citep{krizhevsky2009learning}. 
In terms of training, we adopt the SGD with momentum 0.9, weight decay $10^{-4}$, and batch size 128 for all training processes. We train VGG-19 through 164 epochs with an initial learning rate of 0.1, which is decreased by a factor 10 at epochs 81 and 122; and train ResNet-34 through 300 epochs with an initial learning rate of 0.1, which is decreased by a factor 10 at epochs 150 and 250. 
The ratio of poisoned samples in training set, \ie, $R = \frac{N_{poisoned}}{(N_{poisoned}+N_{benign})}$, is set to 0.25. 
All experiments are conducted on one single GeForce GTX 1080 GPU, and the implementation is conducted based on the open source code\footnote{\url{https://github.com/bearpaw/pytorch-classification}}.

\noindent \textbf{Data Preprocessing.} Before adding a backdoor trigger to the benign sample to generate poisoned samples, we conduct standard data augmentation techniques for benign images. Specifically, 4-pixel padding is used before performing random crops of size $32\times32$.

\section{Settings for Transformation-based Defense}
\label{app:sec_defense}

In this section, we illustrate the detailed settings in Section \ref{transdefense} of the main manuscript.

\noindent \textbf{Defense Setup.}
We examine Flip and ShrinkPad with shrinking size $\in \{1,2,3,4\}$. Except for aforementioned Flip and ShrinkPad, we also conduct fine-pruning \citep{liu2018fine}, neural cleanse \citep{wangneural}, and auto-encoder based defense (dubbed Auto-Encoder) \citep{liu2017neural}, which are the state-of-the-art defenses. The model with standard training and testing process is also provided, which is dubbed {\it Standard}. Specifically, the fine-pruning method consists of two stages, including pruning and fine-tuning. Per the settings in the original paper, we prune the parameters of the last component (convolutional layer for VGG, convolutional block for ResNet). 
The original test set is equally divided as two disjoint subsets, including the validation set and the practical test set. The fraction of pruned neurons is determined through grid-search on the validation set, and the performance is evaluated on the practical test set. In particular, we found that the fine-tuning with even one epoch may reactivate the removed backdoor, therefore it is removed in the experiments. 
For neural cleanse, all settings are based on the open-source code\footnote{\url{https://github.com/bolunwang/backdoor}} provided by the authors.
For Auto-Encoder, we train the convolutional auto-encoder \citep{geng2015} with 100 epochs, learning rate 0.001 and batch size 16. The implementation is based on the open-source code\footnote{\url{https://github.com/jellycsc/PyTorch-CIFAR-10-autoencoder}}. Above defense experiments are conducted on one single GeForce GTX 1080 GPU.

\begin{figure}[ht]
 \centering
 \includegraphics[width=0.8\textwidth]{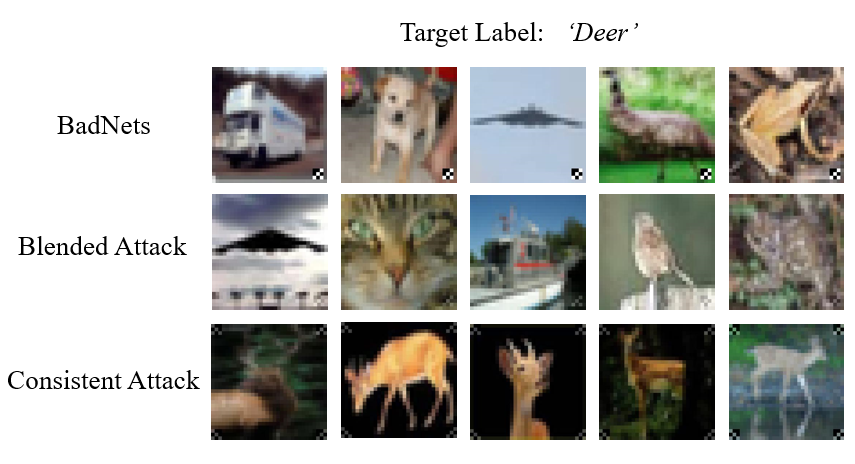}
 \caption{Some poisoned samples generated by different backdoor attack methods. In this experiment, the target label is 'Deer'. Except for the Consistent Attack, the ground-truth label of generated poisoned samples and the target label is not consistent.}
 \label{poiexp}
\end{figure}

\noindent \textbf{Attack Setup.} We use three representative state-of-the-art backdoor attacks, including BadNets \citep{gu2017badnets}, attack with blended strategy \citep{chen2017targeted} (dubbed Blended Attack), and label consistent backdoor attack \citep{turner2019label} (dubbed Consistent Attack) to evaluate the performance of backdoor defenses. The target label is {\it Deer}. Specifically, for BadNets, except for the trigger appearance, other settings are the same as those illustrated in Section \ref{sec_char}. The non-zero pixel value is modified from 128 to 255; For Blended Attack, the trigger is the same as the one of BadNets, the ratio of poisoned samples is set to 0.2, and the hyper-parameter $\bm{\alpha} \in \{0,0.2\}^{3\times32\times32}$. The values of the $\bm{\alpha}$ entries corresponding to the pixels located in the minimum covering box are 0.2, while other values are 0; For Consistent Attack, the ratio of poisoned sample over all training samples with target label is set to 0.25, and $\bm{\alpha} \in \{0,0.25\}^{3\times32\times32}$. The trigger of Consistent Attack is quite different from the one used in BadNets and Blended Attack, which is symmetrical. All these settings follow their original papers. Some examples of poisoned sample generated by different attacks are shown in Figure \ref{poiexp}.

\noindent \textbf {Training Setup.} The training settings are the same as those adopted in Section \ref{sec_char}.

\section{Settings for Attack Enhancement}
\label{app:setattack}
In this section, we illustrate the detailed settings in Section \ref{sec:attack_enhance} of the main manuscript.

\noindent \textbf{Settings}. 
In the enhanced backdoor attacks, we adopt random Flip followed by random ShrinkPad in the random transformation layer. Note that there is only one hyper-parameter in enhanced attacks, $i.e.$, the maximal shrinking size, which is set to 4 pixels in this experiment. We examine three enhanced backdoor attacks, including enhanced BadNets (BadNet+), enhanced Blended Attack (Blended Attack+), and enhanced Consistent Attack (Consistent Attack+) with their correspondingly standard attack in the experiments. In particular, when evaluating the ASR of enhanced attacks under defenses, the random transformation is also adopted on the benign training samples rather than only on the poisoned samples during the training process. This modification is to exclude the possibility that the transformation itself creates a new backdoor. For example, the zero-padding in ShrinkPad may create a new backdoor activated by the black edges of the image. If the random transformations are only adopted on the poisoned samples, we cannot identify whether the improvement of ASR under ShrinkPad is due to that the enhanced attacks are more robust to transformation, or due to that the black edges introduced by ShrinkPad activate the new edge-related backdoor of enhanced attacks. Other settings are the same as those used in Section \ref{transdefense} of the main manuscript.

\section{Defense with Non-spatial Transformation}
\label{defense:non-spatial}

In this section, we examine the effectiveness of proposed transformation-based defense with non-spatial transformations. Specifically, we evaluate two most widely used transformations, including the additive Gaussian noise and the color-shifting, which only change the trigger appearance while preserving its location. 

\begin{figure}[ht]
 \centering
 \includegraphics[width=0.6\textwidth]{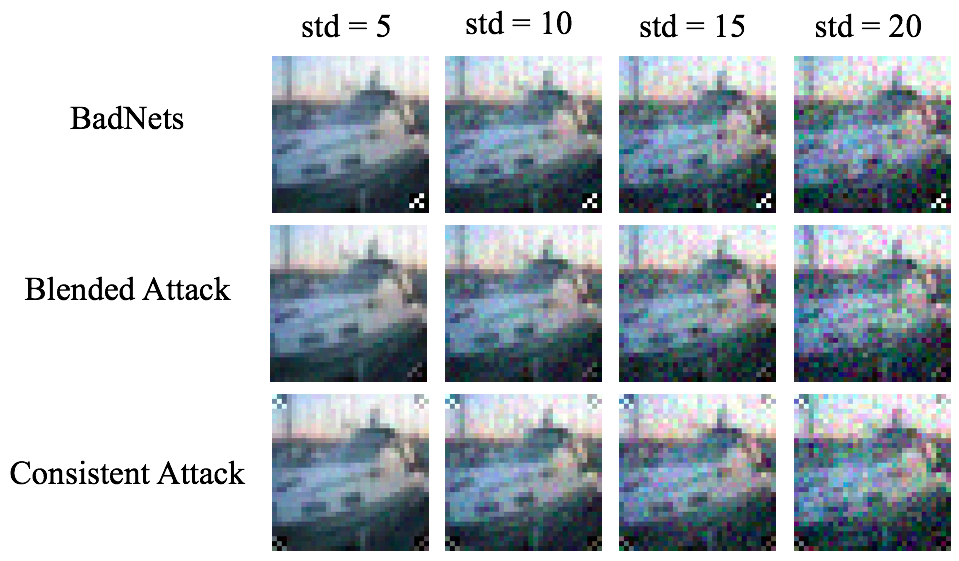}
 \caption{The example of some generated attacked samples with additive Gaussian noise.}
 \label{fig_Gaussian}
\end{figure}

\begin{figure*}[ht]
\centering
\subfigure[Hue]{
\includegraphics[width=0.483\textwidth]{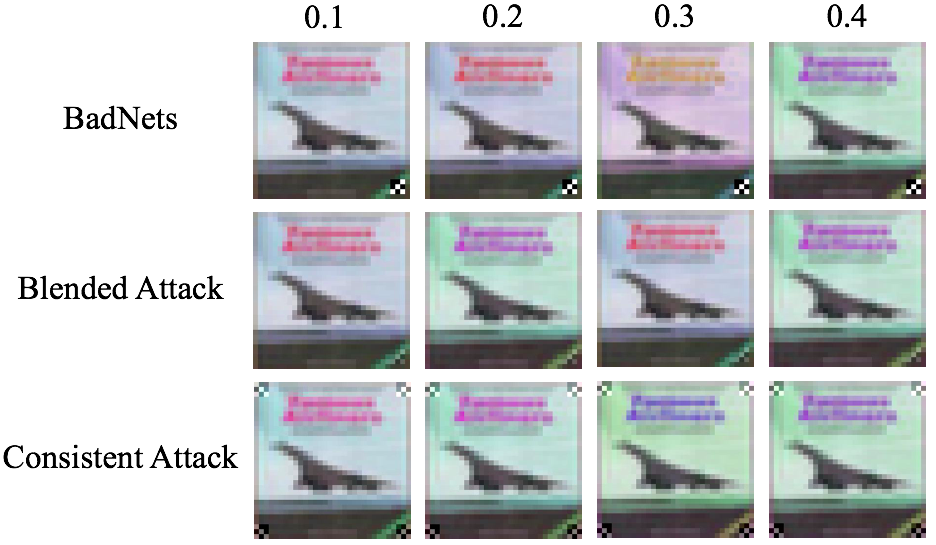}}
\subfigure[Contrast]{
\includegraphics[width=0.483\textwidth]{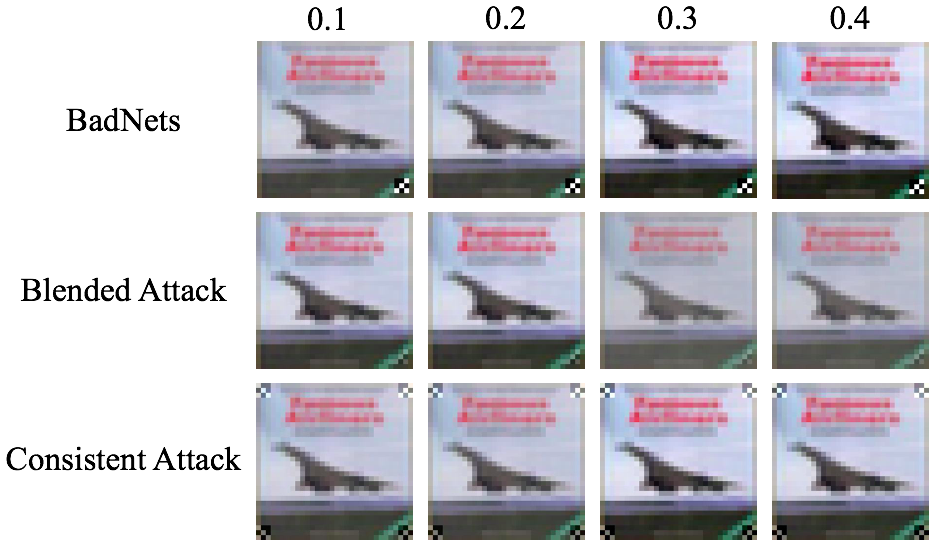}}
\subfigure[Brightness]{
\includegraphics[width=0.483\textwidth]{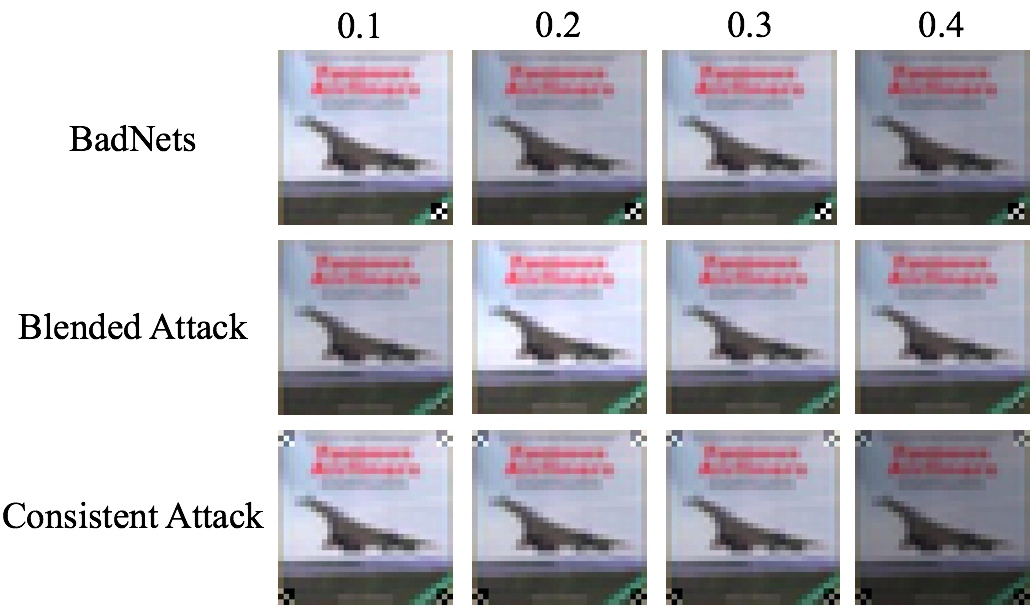}}
\subfigure[Saturation]{
\includegraphics[width=0.483\textwidth]{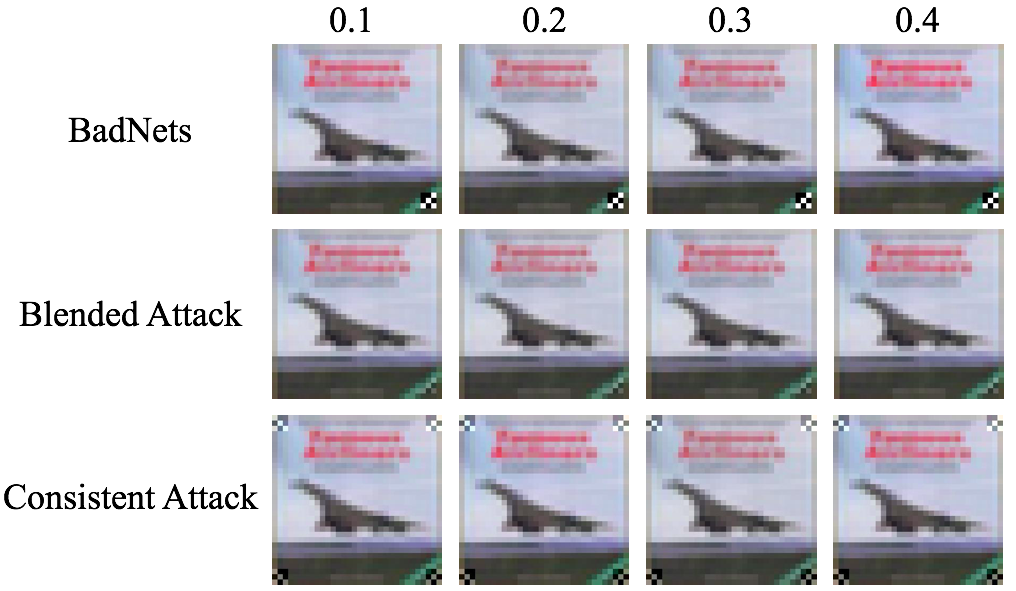}}
\caption{Transformed attacked samples with different types of color-shifting. All images are randomly transformed with maximum perturbation size $\in\{0.1, 0.2, 0.3, 0.4\}$. }
\label{fig_shifting}
\end{figure*}

\noindent \textbf{Settings}. We examine the performance of defense under ResNet-34 structure. For the additive Gaussian noise, the mean is set as zero, and the standard deviation (std), is selected from $\{\frac{5}{255}, \frac{10}{255}, \frac{15}{255}, \frac{20}{255}\}$.
We examine four types of color-shifting, including modifying hue (dubbed Hue), modifying contrast (dubbed Contrast), modifying brightness (dubbed Brightness), and modifying saturation (dubbed Saturation). All images are randomly transformed with maximum perturbation size $\in\{0.1, 0.2, 0.3, 0.4\}$, based on the {\it ColorJitter} function provided in torchvision. Some examples of transformed attacked images are shown in Figures \ref{fig_Gaussian}-\ref{fig_shifting}.

\begin{table*}[ht]
\centering
\small
\caption{Attack success rate and clean test accuracy under additive Gaussian noise with different standard deviation.}
\begin{tabular}{c|cc|cc|cc|cc}
\hline
Standard Deviation (std) $\rightarrow$                  & \multicolumn{2}{c|}{5} & \multicolumn{2}{c|}{10} & \multicolumn{2}{c|}{15} & \multicolumn{2}{c}{20} \\ 
                  \hline
Attack Type $\downarrow$                  & Clean                 & ASR                  & Clean                  & ASR                  & Clean                  & ASR                  & Clean                  & ASR                  \\ \hline
BadNets           & 91.2                  & 100                  & 79.8                   & 100                  & 58.1                   & 100                  & 36.4                   & 100                  \\
Blended Attack    & 90.8                  & 100                  & 81.4                   & 100                  & 64.5                   & 99.9                 & 46.0                   & 99.5                 \\
Consistent Attack & 90.9                  & 98.7                 & 81.9                   & 99.1                 & 65.1                   & 99.4                 & 44.6                   & 99.6                 \\ \hline
\end{tabular}
\label{table_Gaussian}
\end{table*}

\begin{table*}[!ht]
\centering
\scriptsize
\caption{Attack success rate and clean test accuracy under different types of color-shifting with different maximum perturbation sizes. We examine four types of color-shifting, including modifying hue (dubbed Hue), modifying contrast (dubbed Contrast), modifying brightness (dubbed Brightness), and modifying saturation (dubbed Saturation). All images are randomly transformed with maximum perturbation size $\in\{0.1, 0.2, 0.3, 0.4\}$.} 
\begin{tabular}{c|c|cc|cc|cc|cc}
\hline
                            &  Maximum Perturbation Size $\rightarrow$  & \multicolumn{2}{c|}{0.1} & \multicolumn{2}{c|}{0.2} & \multicolumn{2}{c|}{0.3} & \multicolumn{2}{c}{0.4} \\ \hline
Shifting Type $\downarrow$                       & Attack Type $\downarrow$          & Clean       & ASR        & Clean       & ASR        & Clean       & ASR        & Clean       & ASR       \\ \hline
\multirow{3}{*}{Hue}        & BadNets           & 93.2        & 100        & 91.6        & 100        & 89.4        & 100        & 88.5        & 100       \\
                            & Blended Attack    & 92.1        & 100        & 89.8        & 100        & 88          & 100        & 86.9        & 100       \\
                            & Consistent Attack & 91.9        & 98.7       & 89.2        & 98.8       & 87.2        & 99         & 85.8        & 99.1      \\ \hline
\multirow{3}{*}{Contrast}   & BadNets           & 94.2        & 100        & 94.0        & 100        & 93.8        & 100        & 93.7        & 100       \\
                            & Blended Attack    & 92.9        & 100        & 92.9        & 100        & 92.8        & 100        & 92.6        & 100       \\
                            & Consistent Attack & 93.0        & 98.5       & 92.8        & 97.9       & 92.6        & 97.5       & 92.4        & 96.4      \\ \hline
\multirow{3}{*}{Brightness} & BadNets           & 94.1        & 100        & 93.9        & 100        & 93.7        & 100        & 93.4        & 100       \\
                            & Blended Attack    & 93.0        & 100        & 92.9        & 99.8       & 92.7        & 99.0       & 92.4        & 98.4      \\
                            & Consistent Attack & 93.0        & 98.1       & 92.9        & 96.4       & 92.6        & 94.5       & 91.9        & 92.6      \\ \hline
\multirow{3}{*}{Saturation} & BadNets           & 94.1        & 100        & 94.1        & 100        & 94.1        & 100        & 94.0        & 100       \\
                            & Blended Attack    & 93.1        & 100        & 93.1        & 100        & 93.0        & 100        & 93.0        & 100       \\
                            & Consistent Attack & 93.0        & 98.7       & 93.0        & 98.8       & 93.0        & 98.7       & 92.8        & 98.7      \\ \hline
\end{tabular}
\label{table_shifting}
\end{table*}

\noindent \textbf{Results}.
As shown in Tables \ref{table_Gaussian}-\ref{table_shifting}, both the additive Gaussian noise and color-shifting have limited effects on defending backdoor attacks. Especially for the additive Gaussian noise, despite the use of a large standard deviation, ASR has not decreased even though the clean accuracy has decreased by more than $30\%$. Besides, color-shifting has limited effects on both defense performance and clean accuracy. The possible reason is that the effects of these transformations on the trigger appearance are not significant, as shown in Figures \ref{fig_shifting}. Moreover, the exact impact of the difference in trigger appearance on the attack success rate of backdoor attacks are still unclear, which will be further studied in the future work. Accordingly, in the proposed transformation-based defense, we recommend to use spatial-transformations instead of non-spatial transformations. The spatial-transformations may probably change the location and appearance of the trigger simultaneously, and the location has a direct connection to the backdoor activation.

\begin{table}[ht]
\center
\small
\caption{The average training time (seconds) of different defenses. }
\label{time}
\begin{tabular}{@{}ccc@{}}
\toprule
               & VGG-19    & ResNet-34    \\ \midrule
Fine-Pruning   & $\sim$ 400       & $\sim$ 600            \\
Neural Cleanse & $\sim$ 30000         & $\sim$ 80000            \\
Auto-Encoder   & \multicolumn{2}{c}{$\sim$ 2000} \\ \midrule
Flip           & $\sim$ \textbf{0}         & $\sim$ \textbf{0}            \\
ShrinkPad      & $\sim$ \textbf{0}         & $\sim$ \textbf{0}            \\ \bottomrule
\end{tabular}
\end{table}

\newpage
\section{More Results about Transformation-based Defense}
\label{app:moredefense}
\subsection{Comparison of Different Defenses from the Aspect of Efficiency}

The proposed transformation-based defense method only involves an extra simple transformation of the image in the inference process. However, all other baseline methods require additional training or optimization. Compared to state-of-the-art methods, the proposed transformation-based defense requires less additional costs. To verify the efficiency of the proposed method, we report the average training time over defending all three attacks of each defense method, as shown in Table \ref{time}. Besides, there is only one hyper-parameter in ShrinkPad, \ie, the shrinking size, while there are multiple hyper-parameters in compared baseline methods. In conclusion, transformation-based defenses (with spatial transformation) reach competitive performance compared with state-of-the-art defense methods, while with almost no additional computational cost and fewer hyper-parameters to adjust.

\begin{table}[ht]
\small
\centering
\vspace{-1.2em}
\caption{Comparison of different backdoor defenses against BadNets with universal perturbation as the backdoor trigger on CIFAR-10 dataset. `Clean' and `ASR' indicates the accuracy (\%) and attack success rate (\%) on testing set, respectively.}
\begin{tabular}{ccccc}
\toprule
               & \multicolumn{2}{c}{VGG-19} & \multicolumn{2}{c}{ResNet-34} \\ \midrule
               & Clean        & ASR         & Clean          & ASR          \\ \midrule
Standard       & 89.9         & 100         & 93.9           & 100          \\ \midrule
Fine-Pruning   & 46.9         & 0           & 87.8           & 1.6          \\
Neural-Cleanse & 48.0         & 100         & 73.3           & 100          \\ \midrule
Auto-Encoder   & 84.0         & 11.6        & 85.9           & 4.5          \\
Flip           & 88.9         & 3.6         & 93.6           & 0.7          \\
ShrinkPad-4    & 85.0         & 12.4        & 91.5           & 6.2          \\ \bottomrule
\end{tabular}
\vspace{-0.6em}
\end{table}

\begin{figure}[ht]
 \centering
 \includegraphics[width=0.7\textwidth]{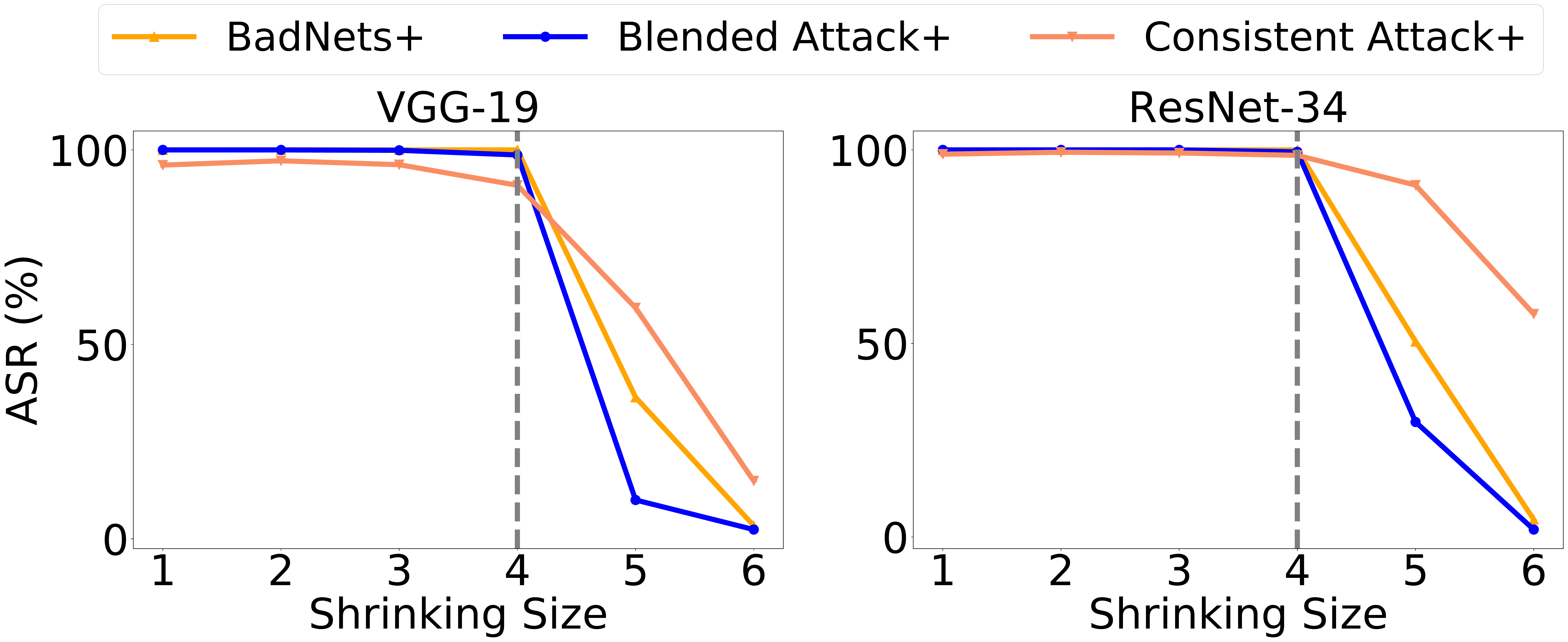}
 \vspace{-0.4em}
 \caption{Attack success rates of enhanced attacks with the maximal shrinking size of 4-pixels, under the ShrinkPad defense with different shrinking sizes.}
 \label{Aba_shrinkpad}
 \vspace{-1em}
\end{figure}

\subsection{Defense against Attack with Universal Perturbation as the Trigger}

Compared with the non-optimized trigger pattern (\eg, a $3\times3$ square, as the one we used in Section \ref{transdefense}), recently, the universal adversarial perturbation \citep{moosavi2017universal} was verified to be a more effective trigger in BadNets-type attack \citep{zhao2020clean}. In this section, we compare different defense methods against a more challenging attack setting, $i.e.$, BadNets with universal perturbation as the backdoor trigger.

\textbf{Settings. } As mentioned above, we adopt the universal perturbation as the backdoor trigger in this experiment. The perturbation is generated based on a pre-trained benign model. The training scheme of the benign model is the same as that of its correspondingly infected version (except for the training set). Other settings are the same as those used in Section \ref{transdefense}.

\textbf{Results. } Compared with the state-of-the-art preprocessing based method ($i.e.,$ Auto-Encoder), our method still has higher clean accuracy and lower ASR under this attack setting. It is interesting to find that both Fine-Pruning and Neural-Cleanse fail in this scenario. For Fine-Pruning, its effectiveness relies on the assumption that the backdoor is hidden in neurons that are not related to those used for encoding the normal behavior of the model. This assumption is not necessarily true in this scenario, since the adversarial perturbation is generated according to the effects of all neurons. The failure of Neural-Cleanse may probably due to the failure of its trigger-reconstruction stage, since the reconstruction of a `noise' is significantly more difficult than that of a compact object (\ie, $3\times3$ square that we used in Section \ref{transdefense}). This experiment again verifies the effectiveness of the proposed method and the importance of designing defenses by utilizing the properties of attacks.

\begin{figure}[ht]
 \centering
 \includegraphics[width=0.6\textwidth]{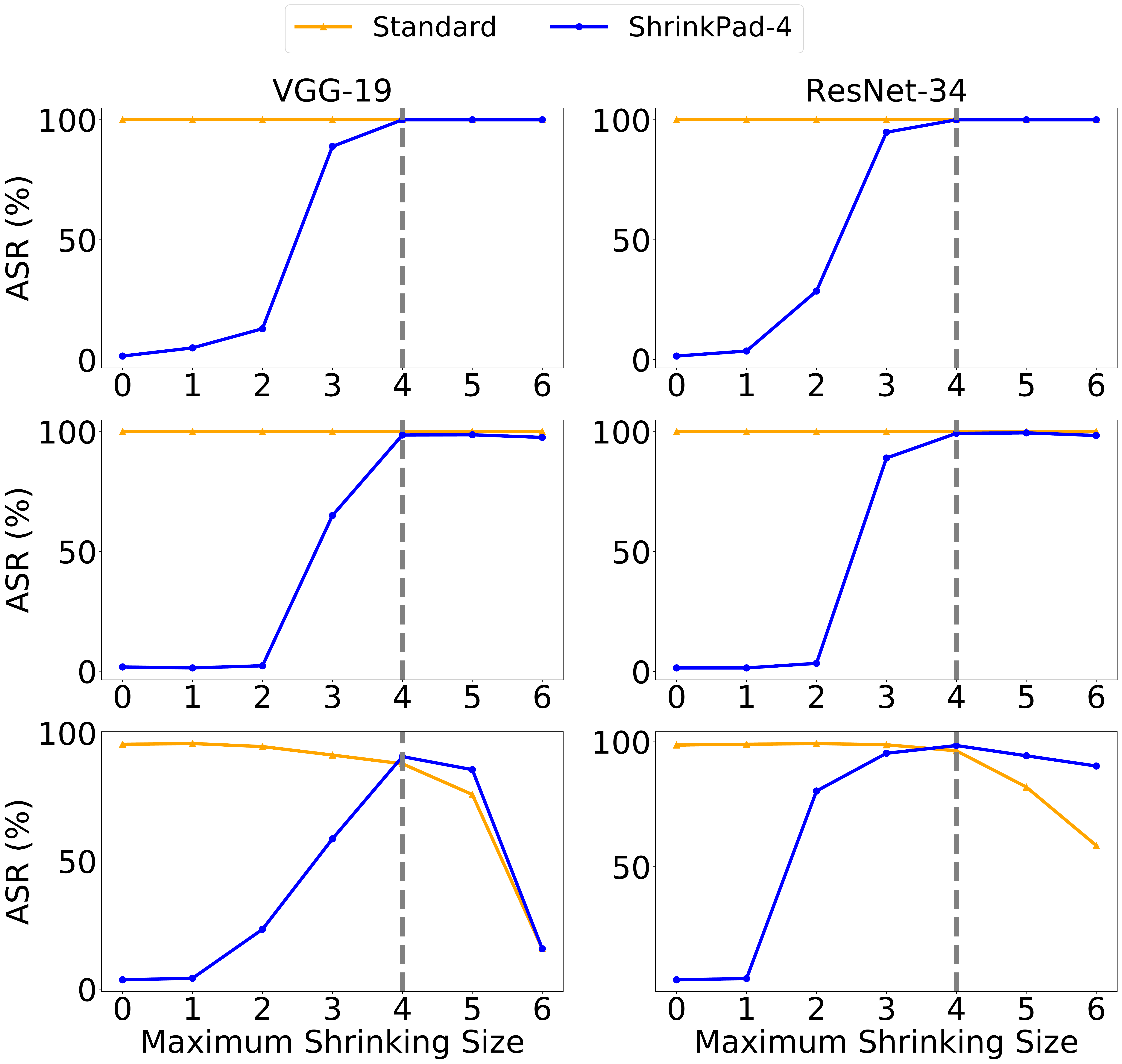}
 \caption{Attack success rate of enhanced backdoor attacks $w.r.t.$ different maximal shrinking sizes under ShrinkPad-4 and Standard. \textbf{First Row:} `BadNets+'; \textbf{Second Row:} `Blended Attack+'; \textbf{Last Row:} `Consistent Attack+'.}
\label{Aba_EnhancedAttack}
\vspace{-1em}
\end{figure}

\vspace{-0.6em}
\section{Ablation Study}
\vspace{-0.5em}
\label{app:aba}

In this section, we study the effect of shrinking size in the proposed defense and the effect of maximal shrinking size in enhanced backdoor attacks. Except for the studied hyper-parameters, other settings are the same as those used in Section \ref{transdefense} and Section \ref{sec:attack_enhance}, unless otherwise specified.

\vspace{-0.4em}
\subsection{The Effect of Shrinking Size in the Transformation-based Defense.}
\vspace{-0.3em}

As demonstrated in Section \ref{transdefense}, adopting ShrinkPad with a small pixels shrinking size will significantly reduce the ASR of standard attacks. In this section, we discuss the effect of shrinking size in defending against enhanced backdoor attacks. Specifically, we evaluate the effect of defending the enhanced attacks with 4-pixels maximal shrinking size.

As shown in Figure \ref{Aba_shrinkpad}, the ASR decreases along with the increase of the shrinking size under all settings. Although when the shrinking size in ShrinkPad is not larger than the maximal shrinking size used in enhanced attacks (\ie, 4 pixels), the ASR values are still very high, indicating that the defense performance of ShrinkPad is not satisfied. However, when the shrinking size is bigger than the maximal shrinking size used in enhanced attacks (4 pixels), the ASR will decrease dramatically. The above results indicate that the shrinking size used in the ShrinkPad defense should be larger than the maximal shrinking size used in enhanced attacks, to ensure the satisfied defense performance.

\subsection{The Effect of Maximal Shrinking Size in Enhanced Backdoor Attacks.}
We evaluate the performance of the enhanced backdoor attack with different maximal shrinking sizes, to attack the Standard model (no defense) and the model with the `ShrinkPad-4' defense. 

The attack results measured by ASR are shown in Figure \ref{Aba_EnhancedAttack}. To attack the Standard model, the ASR values are very high and are almost unchanged when the maximal shrinking size varies. 
However, the ASR values of the Consistent Attack+ decreases along with the increase of the maximal shrinking size. The larger value of the maximal shrinking size indicates the more randomness of triggers in training, which requires more poisoned training images to create the backdoor. As mentioned in Section \ref{sec:attack_enhance} in the main manuscript, the number of poisoned training images in Consistent Attack+ is insufficient. 
To attack the model with the defense ShrinkPad-4, when the maximal shrinking size is smaller than the shrinking size 4 in ShrinkPad-4, the ASR values increase from 0 to almost 100. 
When the maximal shrinking size is larger than the shrinking size 4, the ASR values of BadNets+ and Blended Attack+ are still about 100; but, the ASR values of Consistent Attack+ become to decrease, still due to the insufficiency of poisoned training images. These phenomena indicate that the proposed attack enhancement can indeed reduce the transformation vulnerability of existing backdoor attacks.

\begin{figure}[ht]
 \centering
 \includegraphics[width=0.85\textwidth]{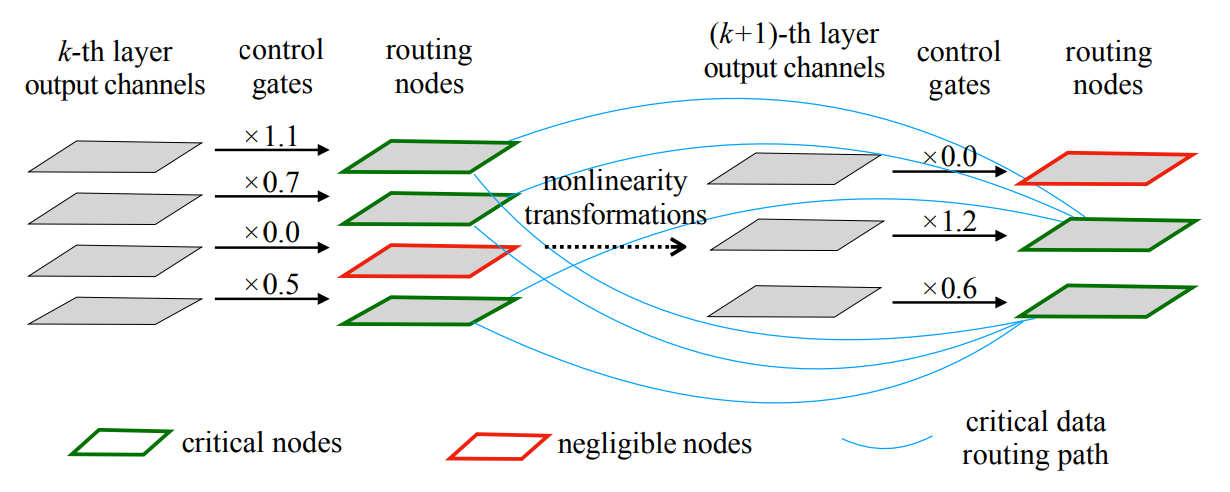}
 \caption{The control gates are multiplied to the layer’s output channel-wise, resulting in the actual routing nodes. The layerwise routing nodes are linked together to compose the routing paths \citep{wang2018cdrp}. }
\label{fig:CDRPs}
\vspace{-1em}
\end{figure}

\section{More Details about the Differences between Standard Backdoor Attack and Enhanced Backdoor Attack}
\label{app:moredetails}

\subsection{A Brief Introduction about Sliency Map and Critical Data Routing Path}

\textbf{Sliency map. }
Saliency map \citep{simonyan2013deep} is widely used in computer vision to provide indications of the most salient regions within images. By creating the saliency map for a DNN model, we can obtain some intuition on \emph{where the network is paying the most attention to} in an input image. 
Specifically, for the image $\bm{x}$ and a classifier with the class score function $S_c(\cdot)$, the image-specific class saliency is the magnitude of the derivative of $S_c(\bm{x})$ $w.r.t.$ $\bm{x}$.

\textbf{Critical data routing path. } The critical data routing paths (CDRPs) \citep{wang2018cdrp} is a distillation guided method, which can be used to interpret DNNs by identifying critical data routing paths and analyzing the functional processing behavior of the corresponding layers. Compared with the sliency map, CDRPs provide more layer-wise information of DNNs. Specifically, it discover the critical nodes on the data routing paths during the inference process for a specific image by learning associated control gates for each layer’s output channel. Accordingly, the routing paths can be represented based on the responses of concatenated control gates from all the layers, which \emph{reflect the network’s semantic selectivity regarding to the input patterns and more detailed functional process across different layer levels}. An illustrative example is shown in Figure \ref{fig:CDRPs}.

Let $f_{\bm{w}}(\cdot)$ be a pretrained network, $f_{\bm{w}}(\cdot;\bm{\Lambda})$ is a network with control gates $\bm{\Lambda} = (\bm{\lambda_1}, \cdots, \bm{\lambda}_K)$ where $\bm{\lambda}_i$ is the control gates of $i$-th layer. The CDRPs of a image $\bm{x}$ is optimized by a distillation-guided method, as follows:

\begin{equation}
\begin{aligned}
& \min_{\bm{\Lambda}} \mathcal{L}(f_{\bm{w}}(\bm{x}), f_{\bm{w}}(\bm{x};\bm{\Lambda})) + \gamma \sum_{i=1}^K |\bm{\lambda}_i|_1  \\
& s.t. \quad \bm{\lambda}_i \succeq 0,\  i = 1, 2, \cdots, K,
\end{aligned}
\end{equation}
where $\mathcal{L}$ is the cross entropy and $\gamma$ is a trade-off hyper-parameter.

\subsection{Settings}
\textbf{Settings for visualizing the sliency map. } 
We visualize the saliency map \citep{simonyan2013deep} towards their predicted label of attacked and transformed attacked images, under both standard attacks and enhanced attacks. The attacked images are transformed by ShrinkPad with 4-pixels shrinking size, and the saliency map is obtained based on the open-source code\footnote{\url{https://github.com/MisaOgura/flashtorch}}.  

\textbf{Settings for visualizing the critical data routing path. } We randomly select 100 benign testing samples (with the ground-truth label different from the target label), and their correspondingly attacked samples and transformed attack samples for generating CDRPs, under the attacks of BadNets and BadNets+. After generating all CDRPs of each sample, we calculate the layerwise correlation coefficients of CDRPs between each sample pair (totally 300 pairs), including (benign sample, attacked sample), (benign sample, transformed attacked sample), and (attacked sample, transformed attacked sample), under BadNets and BadNets+, respectively. The CDRP is obtained based on the open-source code\footnote{\url{https://github.com/frankwang345/cdrp-detect}}.

\end{document}